\documentclass{IEEEtran}
\usepackage{amssymb}
\usepackage{}
\usepackage{amsfonts}
\usepackage{amsmath}
\usepackage{algorithm}
\usepackage{algorithmic}
\usepackage{cite}

\usepackage{graphicx,subfigure,amsmath,amssymb,cite}

\usepackage[dvips]{color}
\usepackage{float}
\ifCLASSINFOpdf
\else
\fi
\begin{document}
\title{Secure SWIPT for Directional Modulation Aided AF Relaying Networks}

\author{\IEEEauthorblockN{Xiaobo Zhou, Jun Li, Feng Shu, Qingqing Wu, Yongpeng Wu, Wen Chen, and Hanzo Lajos}\\
\thanks{Xiaobo Zhou, Jun Li, and Feng Shu are with the School of Electronic and Optical Engineering, Nanjing University of Science and Technology, Nanjing, 210094, China. e-mail: \{zxb,jun.li\}@njust.edu.cn, shufeng0101@163.com. Xiaobo Zhou is also with the School of Fuyang normal university, Fuyang, 236037, China. Qingqing Wu is with the Department of Electrical and Computer Engineering, National University of Singapore, 117583, Singapore. e-mail: elewuqq@nus.edu.sg. Yongpeng Wu and Wen Chen are with Shanghai Jiao Tong University, Minhang 200240, China. e-mail: \{yongpeng.wu,wenchen\}@sjtu.edu.cn. Hanzo Lajos is with the Department of Electronics and Computer Science, University of Southampton, U. K. e-mail:lh@ecs.soton.ac.uk.}}

\maketitle

\begin{abstract}
Secure wireless information and power transfer based on directional modulation is conceived for amplify-and-forward (AF) relaying networks. Explicitly, we first formulate a secrecy rate maximization (SRM) problem, which can be decomposed into a twin-level optimization problem and solved by a one-dimensional (1D) search and semidefinite relaxation (SDR) technique. Then in order to reduce the search complexity, we formulate an optimization problem based on maximizing the signal-to-leakage-AN-noise-ratio (Max-SLANR) criterion, and transform it into a SDR problem. Additionally, the relaxation is proved to be tight according to the classic Karush-Kuhn-Tucker (KKT) conditions. Finally, to reduce the computational complexity, a successive convex approximation (SCA) scheme is proposed to find a near-optimal solution. The complexity of the SCA scheme is much lower than that of the SRM and the Max-SLANR schemes. Simulation results demonstrate that the performance of the SCA scheme is very close to that of the SRM scheme in terms of its secrecy rate and bit error rate (BER), but much better than that of the zero forcing (ZF) scheme.

\end{abstract}
\IEEEpeerreviewmaketitle
\section{Introduction}
The past decade has witnessed the rapid development of Internet of Things (IoT). It is forecast that by 2025 about 30 billion IoT devices will be used worldwide\cite{Wu2018Spectral,Wu2016Joint}. As conventional battery is not convenient for such a huge number of devices, simultaneous wireless information and power transfer (SWIPT) is recognized as a promising technology to prolong the operation time of wireless devices \cite{HSonjoint2014,Chu2016,Zhang2016,Ng2013,Wu2015,Wu2016En,Khandaker2015,ArtificialMEI2017,Xun2013,Zhang2013MIMO,WU2017,Wu2016Energy}. The separated information receiver (IR) and energy receiver (ER) are considered in \cite{HSonjoint2014,Chu2016,Zhang2016}. The authors in \cite{HSonjoint2014,Wu2016User} considered a multi-user wireless information and power transfer system, where the beamforming vector was designed by the zero-forcing (ZF) algorithm and updated by maximizing the energy harvested. In \cite{Chu2016}, the optimal beamforming scheme was proposed for achieving the maximum secrecy rate, while meeting the minimum energy requirement at the ER. In \cite{Zhang2016,Wu2017Energy}, the authors designed the robust information and energy beamforming vectors for maximizing the energy harvested by the ER under specific constraints on the signal-to-interference plus noise ratio (SINR) at the IR. A power splitting (PS) scheme was utilized to divide the received signal into two parts in order to simultaneously harvest energy and to decode information \cite{Ng2013,Wu2015,Khandaker2015,ArtificialMEI2017}. The authors of \cite{Xun2013,Zhang2013MIMO} investigated both PS and time switching (TS) schemes and compared the performance of these two schemes.

As an important technique of expanding the coverage of networks, relaying can also beneficially enhance the communication security, whilst simultaneously enhancing energy harvesting \cite{Chen2015Secrecy,Zou2015A,Zou2014Relay,RelayBC,KimNon}. For the case of perfect channel state information (CSI) situations, secure SWIPT invoked in relaying networks has been investigated \cite{Li2014Secure,Salem2016Physical,Li2016Secure}. Literature \cite{Li2014Secure}, proposed a constrained concave convex procedure (CCCP)-based iterative algorithm for designing the beamforming vector of multi-antenna aided non-regenerative relay networks. While in \cite{Salem2016Physical}, the analytical expressions of the ergodic secrecy capacity were derived separately based on TS, PS and on ideal relaying protocols. The beamforming vectors of SWIPT were designed for amplify and forward (AF) two-way relay networks through a sequential parametric convex approximation (SPCA)-based iterative algorithm to find its locally optimal solution in \cite{Li2016Secure}.

By contrast, for the imperfect CSI scenarios, the channel estimation uncertainty model was considered \cite{Xing2015To,Feng2017Robust,Li2017Secure,Niu2017Joint,Li2016Robust}. In \cite{Xing2015To,Feng2017Robust,Li2017Secure}, the robust information beamformer and artificial noise (AN) covariance matrix were designed with the objective of maximizing the secrecy rate under the constraint of a certain maximum transmit power. The secrecy rate maximization (SRM) problem was a non-convex problem in \cite{Xing2015To,Feng2017Robust,Li2017Secure}, the critical process is, how to transform it into a tractable convex optimization problem by using the S-Procedure. In \cite{Niu2017Joint} and \cite{Li2016Robust}, the authors formulate the power minimization problem under a specific secrecy rate constraint and minimum energy requirement at the energy harvester (EH), which was solved in a similar manner. In general, for the imperfect CSI situations, the channel estimation error is usually modeled obeying the ellipsoid bound constraint, and then be transformed into a convex constraint by using the S-procedure.
%

Recently, a promising physical layer security technique, known as directional modulation (DM), has attracted a lot of attention. In contrast to conventional information beamforming techniques, DM has the ability to directly transmit the confidential messages in desired directions to guarantee the security of information transmission, while distorting the signals leaking out in other directions\cite{MP2009Directional,Yuan2014A,Ding2015Orthogonal,Hu2016Robust,Shu2016Robust}. In \cite{MP2009Directional}, the authors proposed a DM technique that employed a phased array to generate the modulation. By controlling the phase shift for each array element, the magnitude and phase of each symbol can be adjusted in the desired direction. The authors in \cite{Yuan2014A} proposed a method of orthogonal vectors and introduced the concept of AN into DM systems and synthesis. Since the AN contaminates the undesired receiver, the security of the DM systems is greatly improved. Subsequently, the orthogonal vector method was applied to the synthesis of multi-beam DM systems \cite{Ding2015Orthogonal}. The proposed methods in \cite{Yuan2014A} and \cite{Ding2015Orthogonal} achieve better performance at the perfect direction angle, but it is very sensitive to the estimation error of the direction angle. The authors in \cite{Hu2016Robust} modeled the error of angle estimation as uniform distribution and proposed a robust synthesis method for the DM system to reduce the effect of estimation error. In \cite{Shu2016Robust}, the authors also considered the estimation error of the direction angle and proposed a robust beamforming scheme in the DM broadcast scenario.

However, none of these contributions consider DM-based relaying techniques. For example, if the desired user is beyond the coverage of the transmitter or there is no direct link between the transmitter and the desired user, the above methods are not applicable. Moreover, in \cite{Hu2016Robust,Shu2016Robust}, the proposed robust methods only designed the normalized confidential messages beamforming and AN projection matrix without considering the power allocation problem. In fact, the power allocation of confidential messages and AN has a great impact on the security of DM systems. To the best of our knowledge, there exists no DM-based scheme considering secure SWIPT, which thus motivates this work.


To tackle this open problem, we propose a secure SWIPT scheme based on AF aided DM. Compared to \cite{Xing2015To,Feng2017Robust,Li2017Secure,Niu2017Joint,Li2016Robust}, instead of channel estimation error modeled obeying the ellipsoid bound constraint, we model the estimation error of direction angle as the truncated Gaussian distribution which is more practical in our DM scenario \cite{Shu2016Robust}. {\bf{The main contributions of this paper are summarized as follows}}.

1) We formulate the SRM problem subject to the total power constraint at an AF relay and to the minimum energy requirement at the ER. Since the secrecy rate expression is the difference of two logarithmic functions, it is noncovex and difficult to tackle directly. Additionally, the estimates of the eavesdropper directions are usually biased. To solve this problem and to find                                                                                                                               the robust information beamforming matrix as well as the AN covariance matrix, we convert the original problem into a twin-level optimization problem, which can be solved by a one-dimensional (1D) search and the classic semidefinite relaxation (SDR) technique. The 1D search range is bounded into a feasible interval. Furthermore, the SDR is proved tight by invoking the Karush-Kuhn-Tucker (KKT) condition.

2) To reduce the the search complexity, we propose a suboptimal solution for maximizing the signal-to-leakage-AN-noise-ratio (Max-SLANR) subject to the total power constraint of the relay and to the minimum energy required at the ER. Due to the existence of multiple eavesdroppers, we consider the sum-power of the confidential messages leaked out to all the eavesdroppers. This optimization problem is also shown to be nonconvex, but it can be transformed into a semidefinite programming (SDP) problem and then solved by the SDR technique. Its tightness is also quantified. To further reduce the computational complexity, we propose an algorithm based on successive convex approximation (SCA). Specifically, we first formulate the SRM optimization problem and then transform it into a second-order cone programming (SOCP) which is finally solved by the SCA method. Furthermore, we analyse and compare the complexity of the aforementioned three schemes.

3) The formulated optimization problems include random variables corresponding to the estimation error of the direction angles, which makes the optimization problems very difficult to tackle directly. To facilitate solving this problem, we derive the analytical expression of the covariance matrix of each eavesdroppers' steering vector and substitute it into the optimization problems to replace the random variable. Moveover, we add relay and energy harvesting node to the DM-based secure systems, which further expand the application of DM technology. Simulation results demonstrate that the bit error rate (BER) performance of all our schemes in the desired direction is significantly better than that in other directions, while the BER is poor in the vicinity of the eavesdroppers' directions, showing the advantages of our DM technology in the field of physical layer security.


The rest of this paper is organized as follows. Section \ref{sec_sys} introduces the system model. In Section \ref{sec_rob}, three algorithms are proposed to design the robust secure beamforming. Section \ref{sec_simu} provides our simulation results. while, Section \ref{conclusion} concludes the paper.

$\emph{Notation}$: Boldface lowercase and uppercase letters represent vectors and matrices, respectively, $\mathbf{A}^*$, $\mathbf{A}^T$, $\mathbf{A}^H$, $\mathrm{Tr}(\mathrm{\mathbf{A}})$, $\mathrm{rank}(\mathrm{\mathbf{A}})$, $\mathbf{A}\succ \mathbf{0}$ and $\mathbf{A}\succeq \mathbf{0}$ denote conjugate, transpose, conjugate transpose, trace, rank, positiveness and semidefiniteness of matrix $\mathrm{\mathbf{A}}$, respectively, $\mathbb{E}[\cdot]$, $j$, and $\|\cdot\|$ denote the statistical expectation,  pure imaginary number, and Euclidean norm, respectively, and $\otimes$ denotes the Kronecker product.
\section{System Model}\label{sec_sys}
\begin{figure}[!ht]
  \centering
  \includegraphics[width=0.45\textwidth]{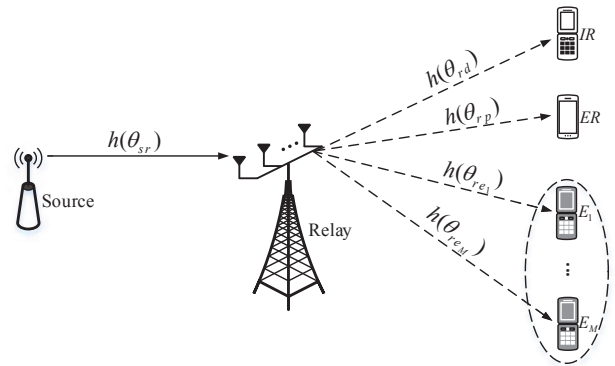}\\
  \caption{System model of secure beamforming with SWIPT based on directional modulation in AF relaying networks}\label{Sys_Sch}
\end{figure}

As shown in Fig.~\ref{Sys_Sch}, we consider AF-aided secure SWIPT, where the source transmitter sends confidential messages to an IR with the aid of an AF relay in the presence of an ER and $M$ eavesdroppers ($\mathrm{E}_1,\cdots, \mathrm{E}_M$). It is assumed that the AF relay is equipped with an $N$-element antenna array, while all other nodes have a single antenna.

Similar to the literature on DM \cite{Hu2016Robust,Shu2016Robust}, this paper adopts the free-space path loss model which is practical for some scenarios such as communication in the air and rural areas. The steering vector between node $a$ and node $b$ can be expressed as \cite{Yuan2014A}
\begin{align}\label{steer}
\mathbf{h}(\theta_{ab})=\sqrt{g_{ab}}\underbrace{\frac{1}{\sqrt{N}}\left[e^{j2\pi\Psi_{\theta_{ab}}(1)}, \cdots, e^{j2\pi\Psi_{\theta_{ab}}(N)}\right]^T}_{\text{The~normalized~steering~vector}},
\end{align}
where $g_{ab}$ is the path loss between node $a$ and node $b$. The function $\Psi_{\theta_{ab}}(n)$ can be expressed as
\begin{equation}\label{var_phi}
\Psi_{\theta_{ab}}(n)\triangleq-\frac{(n-(N+1)/2)l\cos\theta_{ab}}{\lambda},n=1,\ldots,N,
\end{equation}
where $\theta_{ab}$ denotes the angle of direction between node $a$ and node $b$, $l$ denotes the distance between two adjacent antenna elements, and $\lambda$ is the wavelength.

We assume that there is no direct link from the source to the IR, ER or to any of the eavesdroppers. Thus the relay helps the source to transmit the confidential message $x$ to IR. The relay node is assumed to operate in an AF half-duplex mode. Simultaneously, ER intends to harvest energy, while the eavesdroppers try to intercept the confidential message. The power of the signal $x$ is normalized to, $\mathbb{E}[xx^H]=1$. In the first time slot, the source transmits the signal $x$ to the relay, and the signal received at the relay is given by
\begin{align}\label{y_r}
\mathbf{y}_r=\sqrt{P_s}\mathbf{h}(\theta_{sr})x+\mathbf{n}_r,
\end{align}
where $P_s$ is the transmission power of the source, $\mathbf{h}(\theta_{sr})$ denotes the steering vector between the source and the relay, $\mathbf{n}_r\sim\mathcal {C}\mathcal {N}(0, \sigma^2_r\mathbf{I}_N)$ is a circularly symmetric complex Gaussian (CSCG) noise vector, and $\theta_{sr}$ is the angle of direction between the source and the relay.
In the second time slot, the relay amplifies and forwards the received signal to IR. The signal transmitted from the relay is given by
\begin{align}\label{xr}
\mathbf{x}_r=\mathbf{W}\mathbf{y}_r+\mathbf{z}=\sqrt{P_s}\mathbf{W}\mathbf{h}(\theta_{sr})x+\mathbf{W}\mathbf{n}_r+\mathbf{z},
\end{align}
where $\mathbf{W}\in \mathbb{C}^{N\times N}$ is the beamforming matrix, and $\mathbf{z}\in \mathbb{C}^{N\times 1}$ is the AN vector assumed to obey a (CSCG) distribution $\mathcal {C}\mathcal {N}(0, \mathbf{\Omega})$ with $\mathbf{\Omega}\succeq 0$. In general,  the relay has a total transmit power constraint $P_t$, therefore we have
\begin{align}\label{P}
P_s\|\mathbf{W}\mathbf{h}(\theta_{sr})\|^2+\sigma_r^2\mathrm{Tr}(\mathbf{W}^H\mathbf{W})+\mathrm{Tr}(\mathbf{\Omega})\leq P_t.
\end{align}
The signal received at the IR, ER, and the $m$-th eavesdropper can be expressed as
\begin{align}\label{y0}
y_d=&\mathbf{h}^H(\theta_{rd})\mathbf{x}_r+n_0\nonumber\\
=&\sqrt{P_s}\mathbf{h}^H(\theta_{rd})\mathbf{W}\mathbf{h}(\theta_{sr})x+\mathbf{h}^H(\theta_{rd})\mathbf{W}\mathbf{n}_r\nonumber\\
&+\mathbf{h}^H(\theta_{rd})\mathbf{z}+n_0,
\end{align}
\begin{align}\label{yk}
y_p=&\sqrt{P_s}\mathbf{h}^H(\theta_{rp})\mathbf{W}\mathbf{h}(\theta_{sr})x+\mathbf{h}^H(\theta_{rp})\mathbf{W}\mathbf{n}_r\nonumber\\
&+\mathbf{h}^H(\theta_{rp})\mathbf{z}+n_p,
\end{align}
and
\begin{align}\label{ye}
y_{e_m}=&\sqrt{P_s}\mathbf{h}^H(\theta_{re_m})\mathbf{W}\mathbf{h}(\theta_{sr})x+\mathbf{h}^H(\theta_{re_m})\mathbf{W}\mathbf{n}_r\nonumber\\
&+\mathbf{h}^H(\theta_{re_m})\mathbf{z}+n_e, m\in \mathcal{M}=[1,2,...,M],
\end{align}
respectively, where $\mathbf{h}(\theta_{rd})$, $\mathbf{h}(\theta_{rp})$, and $\mathbf{h}(\theta_{re_m})$
denote the steering vectors from the relay to IR, ER, and the $m$-th eavesdropper respectively. Furthermore, $n_0$, $n_p$, and $n_e$ represent the CSCG noise at IR, ER, and the $m$-th eavesdropper, respectively, while $n_0\sim\mathcal {C}\mathcal {N}(0, \sigma_0^2)$, $n_p\sim\mathcal {C}\mathcal {N}(0, \sigma_k^2)$, and $n_e\sim\mathcal {C}\mathcal {N}(0, \sigma_e^2)$. Without loss of generality, we assume that $\sigma_r^2$, $\sigma_0^2$, $\sigma_k^2$, and $\sigma_e^2$ are all equal to $\sigma^2$.

Similar to the considerations in~\cite{Feng2017Robust} and \cite{Niu2017Joint}, namely that the perfect CSI of the destination is available at the relay, here we assume that the relay has the perfect knowledge of direction angles to the IR. However, there is an estimation error of the direction angles of eavesdroppers at the relay, and we assume that the relay has the statistical information about these estimation errors. Therefore, the $m$-th eavesdropper's direction angle to the relay can be modeled as
\begin{align}\label{dir_e}
\theta_{re_m}=\hat{\theta}_{re_m}+\Delta\theta_{re_m}, m\in \mathcal{M},
\end{align}
where $\hat{\theta}_{re_m}$ is the estimate of the $m$-th eavesdropper's direction angle at the relay, and $\Delta\theta_{re_m}$ denotes the estimation error, while is assumed to follow a truncated Gaussian distribution spread over the interval $[-\Delta\theta_{max},\Delta\theta_{max}]$ with zero mean and variance $\sigma_\theta^2$.
The probability density function of $\Delta\theta_{re_m}$ can be expressed as
\begin{align}\label{ZF17}
&f\left(\Delta\theta_{re_m}\right)=\nonumber\\
&\begin{cases}
\frac{1}{K_e\sqrt{2\pi}\sigma_{\theta}}e^{\frac{-\Delta\theta_{re_m}^2}{2\sigma_{\theta}^2}},& -\Delta\theta_{max}\leq\Delta\theta_{re_m}\leq\Delta\theta_{max},\\
0,& \mathrm{otherwise},\\
\end{cases}
\end{align}

where $K_e$ is the normalization factor defined as
\begin{align}\label{Ke}
K_{e}=\int_{-\Delta\theta_{max}}^{\Delta\theta_{max}}\frac{1}{\sqrt{2\pi}\sigma_{\theta}}e^{\frac{-\Delta\theta_{re_m}^2}{2\sigma_{\theta}^2}}d(\Delta\theta_{re_m}).
\end{align}


\section{Robust Secure SWIPT Design}\label{sec_rob}
In this section, three algorithms are proposed to design the robust secure beamforming under the assumption that an estimation error
of the direction angles of eavesdroppers exists at the relay. To design the robust beamforming matrix and AN covariance matrix, we first define
\begin{align}\label{imp1}
\mathbf{H}_{re_m}\triangleq\mathbb{E}\left[\mathbf{h}(\hat{\theta}_{re_m}+\Delta\theta_{re_m})\mathbf{h}^H(\hat{\theta}_{re_m}+\Delta\theta_{re_m})\right],m\in \mathcal{M},
\end{align}
and $\mathbf{H}_{re_m}\in \mathbb{C}^{N \times N}$. Let $H_{re_m}(p,q)$ denote the $p$-th row and $q$-th column entry of $\mathbf{H}_{re_m}$, and $H_{re_m}(p,q)$ can be written as
\begin{align}\label{imp2}
H_{re_m}(p,q)=\Gamma_{1_m}(p,q)-j\Gamma_{2_m}(p,q),
\end{align}
where $\Gamma_{1_m}$ and $\Gamma_{2_m}$ can be found in $(\ref{APP1_8})$ and $(\ref{APP1_11})$, respectively. The specific derivation procedure is detailed in Appendix A.

According to $(\ref{yk})$, the energy harvested at the ER is given by \cite{Liu2014Secrecy}
\begin{align}\label{ER}
E=&\rho\big[P_s|\mathbf{h}^H(\theta_{rp})\mathbf{W}\mathbf{h}(\theta_{sr})|^2+\sigma^2\|\mathbf{W}^H\mathbf{h}(\theta_{rp})\|^2\nonumber\\
&+\mathbf{h}^H(\theta_{rp})\mathbf{\Omega}\mathbf{h}(\theta_{rp})\big],
\end{align}
where $0<\rho\leq 1$ denotes the energy transfer efficiency of the ER.

From $(\ref{y0})$, the SINR at the IR can be expressed as
\begin{align}\label{SINRd}
\mathrm{SINR}_d=\frac{P_s|\mathbf{h}^H(\theta_{rd})\mathbf{W}\mathbf{h}(\theta_{sr})|^2}{\sigma^2\|\mathbf{W}^H\mathbf{h}(\theta_{rd})\|^2+\mathbf{h}^H(\theta_{rd})\mathbf{\Omega}\mathbf{h}(\theta_{rd})+\sigma^2}.
\end{align}

According to $(\ref{ye})$, the $m$-th eavesdropper's $\mathrm{SINR}$ is given by
\begin{align}\label{SINRe}
&\mathrm{SINR}_{e_m}=\frac{P_s|\mathbf{h}^H(\theta_{re_m})\mathbf{W}\mathbf{h}(\theta_{sr})|^2}{\sigma^2\|\mathbf{W}^H\mathbf{h}(\theta_{re_m})\|^2+\mathbf{h}^H(\theta_{re_m})\mathbf{\Omega}\mathbf{h}(\theta_{re_m})+\sigma^2}.
\end{align}

Thus, the achievable secrecy rate at the IR can be expressed as \cite{Improving_Dong}
\begin{align}\label{Rs}
R_s & =\min_{m\in \mathcal{M}}\frac{1}{2}\{\log_2\left(1+\mathrm{SINR}_d\right)-\mathbb{E}\left[\log_2\left(1+\mathrm{SINR}_{e_m}\right)\right]\},
\end{align}
where the scaling factor $\frac{1}{2}$ is due to the fact that two time slots are required to transmit one message. By invoking Jensen's inequality, the worst-case secrecy rate is given by
\begin{align}\label{Rsinq}
&R_s\geq \bar{R}_s=\nonumber\\
&\min_{m\in \mathcal{M}}\frac{1}{2}\{\log_2\left(1+\mathrm{SINR}_d\right)-\log_2\left(1+\mathbb{E}\left[\mathrm{SINR}_{e_m}\right]\right)\},
\end{align}
where the expectation of the $\mathrm{SINR_{em}}$ can be approximated as \cite{Mukherjee2010Robust}\cite{Alam2017Asymptotic}
\begin{align}\label{ESINRe}
&\mathbb{E}\left[\mathrm{SINR}_{e_m}\right]\approx\\
&\frac{\mathbb{E}[P_s|\mathbf{h}^H(\theta_{re_m})\mathbf{W}\mathbf{h}(\theta_{sr})|^2]}{\mathbb{E}[\sigma^2\|\mathbf{W}^H\mathbf{h}(\theta_{re_m})\|^2]+\mathbb{E}[\mathbf{h}^H(\theta_{re_m})\mathbf{\Omega}\mathbf{h}(\theta_{re_m})]+\sigma^2}\nonumber.
\end{align}
\subsection{Secrecy Rate Maximization based on One-Dimensional search Scheme (SRM-1D)}\label{sub_SRM}
In this subsection, the robust information beamforming matrix $\mathbf{W}$ and AN covariance matrix $\mathbf{\Omega}$ are designed by our SRM-1D scheme. Specifically, according to $(\ref{P})$, $(\ref{ER})$, and $(\ref{Rs})$, we maximize the worst-case secrecy rate subject to the total transmit power and the harvested energy constraints. Then the optimization problem can be formulated as
\begin{subequations}\label{P1:1}
\begin{align}
&\mathrm{(P1):}\max_{\mathbf{W},\mathbf{\Omega}}~\bar{R}_s\label{P1:1a}\\
&\mathrm{s.t.}~P_s\|\mathbf{W}\mathbf{h}(\theta_{sr})\|^2+\sigma^2\mathrm{Tr}(\mathbf{W}^H\mathbf{W})+\mathrm{Tr}(\mathbf{\Omega})\leq P_t, \label{P1:1b}\\
&~~~~~\rho\big[P_s|\mathbf{h}^H(\theta_{rp})\mathbf{W}\mathbf{h}(\theta_{sr})|^2+\sigma^2\|\mathbf{W}^H\mathbf{h}(\theta_{rp})\|^2\nonumber\\
&~~~~~+\mathbf{h}^H(\theta_{rp})\mathbf{\Omega}\mathbf{h}(\theta_{rp})\big]\geq P_{min},\mathbf{\Omega}\succeq\mathbf{0}\label{P1:1c},
\end{align}
\end{subequations}
where $(\ref{P1:1b})$ denotes total power constraint at the relay, and the first term in $(\ref{P1:1c})$ denotes the minimum power required by the ER. We employ a 1D search and a SDR-based algorithm to solve problem (P1). Observe that $\bar{R}_s$ is the difference of two logarithmic functions, which is non-convex and untractable. Similar to \cite{Liu2014Secrecy}, we decompose  $(\ref{P1:1})$ into two sub-problems, yielding:
\begin{align}\label{per1}
&\max_{\beta}~\frac{1}{2}\log_2\left(\frac{1+\phi(\beta)}{1+\beta}\right)\nonumber\\
&\mathrm{s.t.}~0\leq\beta\leq\beta_{max},
\end{align}
and
\begin{align}\label{per2}
&\phi(\beta)=\max_{{\mathbf{W},\mathbf{\Omega}}}~ \frac{P_s|\mathbf{h}^H(\theta_{rd})\mathbf{W}\mathbf{h}(\theta_{sr})|^2}{\sigma^2\|\mathbf{W}^H\mathbf{h}(\theta_{rd})\|^2+\mathbf{h}^H(\theta_{rd})\mathbf{\Omega}\mathbf{h}(\theta_{rd})+\sigma^2}\nonumber\\
&\mathrm{s.t.}~\frac{\mathbb{E}[P_s|\mathbf{h}^H(\theta_{re_m})\mathbf{W}\mathbf{h}(\theta_{sr})|^2]}{\mathbb{E}[\sigma^2\|\mathbf{W}^H\mathbf{h}(\theta_{re_m})\|^2]+\mathbb{E}[\mathbf{h}^H(\theta_{re_m})\mathbf{\Omega}\mathbf{h}(\theta_{re_m})]+\sigma^2}\nonumber\\
&~~~~~\leq\beta,m\in \mathcal{M},\nonumber\\
&~~~~~(\ref{P1:1b}),(\ref{P1:1c}),
\end{align}
where $\beta$ is a slack variable.
The main steps to solve the problem (P1) are as follows. First, for each $\beta$ inside the interval $[0,\beta_{max}]$, we can obtain a corresponding $\phi(\beta)$ by solving the problem $(\ref{per2})$. Second, upon substituting $\beta$ and $\phi(\beta)$ into the objective function of $(\ref{per1})$, we obtain the secrecy rate corresponding to the given $\beta$. Thirdly, we perform a 1D search for $\beta$, compare all the secrecy rates obtained and then finally we find the optimal value for $(\ref{per1})$.

As for the above procedure of solving the problem (P1), the most important and complex part is to solve the problem $(\ref{per2})$ to obtain $\phi(\beta)$. This are illustrated as follows. Upon defining $\mathbf{w}\triangleq \mathrm{vec}(\mathbf{W})\in \mathbb{C}^{N^2\times 1}$, we can rewrite $(\ref{per2})$ as
\begin{subequations}\label{per3}
\begin{align}
&\phi(\beta)=\max_{{\mathbf{w},\mathbf{\Omega}}}~ \frac{P_s\mathbf{w}^H\mathbf{A}_1\mathbf{w}}{\sigma^2\mathbf{w}^H\mathbf{A}_2\mathbf{w}+\mathbf{h}^H(\theta_{rd})\mathbf{\Omega}\mathbf{h}(\theta_{rd})+\sigma^2},\label{per3a}\\
&\mathrm{s.t.}~\frac{P_s\mathbf{w}^H\mathbf{B}_{1_m}\mathbf{w}}{\sigma^2\mathbf{w}^H\mathbf{B}_{2_m}\mathbf{w}+\mathrm{Tr}(\mathbf{H}_{re_m}\mathbf{\Omega})+\sigma^2}\leq\beta,m\in \mathcal{M},\label{per3b}\\
&~~~~~P_s\mathbf{w}^H\mathbf{C}_1\mathbf{w}+\sigma^2\mathbf{w}^H\mathbf{w}+\mathrm{Tr}(\mathbf{\Omega})\leq P_t, \\
&~~~~~P_s\mathbf{w}^H\mathbf{D}_1\mathbf{w}+\sigma^2\mathbf{w}^H\mathbf{D}_2\mathbf{w}+\mathbf{h}^H(\theta_{rp})\mathbf{\Omega}\mathbf{h}(\theta_{rp})\nonumber\\
&~~~~~\geq\frac{P_{min}}{\rho},~\mathbf{\Omega}\succeq\mathbf{0},
\end{align}
\end{subequations}
where
\begin{subequations}\label{per4:1}
\begin{align}
\mathbf{A}_1=&\left[\mathbf{h}^*(\theta_{sr})\mathbf{h}^T(\theta_{sr})\right]\otimes \left[\mathbf{h}(\theta_{rd})\mathbf{h}^H(\theta_{rd})\right],\label{per4:1a}\\
\mathbf{A}_2=&\mathbf{I}_N\otimes \left[\mathbf{h}(\theta_{rd})\mathbf{h}^H(\theta_{rd})\right],\\
\mathbf{B}_{1_m}=&\left[\mathbf{h}^*(\theta_{sr})\mathbf{h}^T(\theta_{sr})\right]\otimes \left(\mathbf{H}_{re_m}\right), m\in \mathcal{M},\label{per4:1c}\\
\mathbf{B}_{2_m}=&\mathbf{I}_N\otimes \left(\mathbf{H}_{re_m}\right),\label{per4:1d}m\in \mathcal{M},\\
\mathbf{C}_1=&\left[\mathbf{h}^*(\theta_{sr})\mathbf{h}^T(\theta_{sr})\right]\otimes \mathbf{I}_N,\\
\mathbf{D}_1=&\left[\mathbf{h}^*(\theta_{sr})\mathbf{h}^T(\theta_{sr})\right]\otimes \left[\mathbf{h}(\theta_{rp})\mathbf{h}^H(\theta_{rp})\right],\\
\mathbf{D}_2=&\mathbf{I}_N\otimes \left[\mathbf{h}(\theta_{rp})\mathbf{h}^H(\theta_{rp})\right].
\end{align}
\end{subequations}

With the above vectorization, we show problem $(\ref{per3})$ can be transformed into a standard SDP problem. Upon defining $\tilde{\mathbf{W}}\triangleq \mathbf{w}\mathbf{w}^H\in\mathbb{C}^{N^2\times N^2}$, $(\ref{per3})$ can be rewritten as
\begin{subequations}\label{per7:1}
\begin{align}
&\phi(\beta)=\max_{{\tilde{\mathbf{W}},\mathbf{\Omega}}}~ \frac{P_s\mathrm{Tr}(\mathbf{A}_1\tilde{\mathbf{W}})}{\sigma^2\mathrm{Tr}(\mathbf{A}_2\tilde{\mathbf{W}})+\mathbf{h}^H(\theta_{rd})\mathbf{\Omega}\mathbf{h}(\theta_{rd})+\sigma^2}\label{per7:1a}\\
&\mathrm{s.t.}~P_s\mathrm{Tr}(\mathbf{B}_{1_m}\tilde{\mathbf{W}})-\beta\sigma^2\mathrm{Tr}(\mathbf{B}_{2_m}\tilde{\mathbf{W}})-\beta\mathrm{Tr}(\mathbf{H}_{re_m}\mathbf{\Omega})\nonumber\\
&~~~~~-\beta\sigma^2\leq 0,m\in \mathcal{M},\label{per7:1b}\\
&~~~~~P_s\mathrm{Tr}(\mathbf{C}_1\tilde{\mathbf{W}})+\sigma^2\mathrm{Tr}(\tilde{\mathbf{W}})+\mathrm{Tr}(\mathbf{\Omega})\leq P_t, \label{per7:1c}\\
&~~~~~P_s\mathrm{Tr}(\mathbf{D}_1\tilde{\mathbf{W}})+\sigma^2\mathrm{Tr}(\mathbf{D}_2\tilde{\mathbf{W}})+\mathbf{h}^H(\theta_{rp})\mathbf{\Omega}\mathbf{h}(\theta_{rp})\nonumber\\
&~~~~~\geq \frac{P_{min}}{\rho},\label{per7:1d}\\
&~~~~~\mathrm{rank}(\tilde{\mathbf{W}})=1,\tilde{\mathbf{W}}\succeq\mathbf{0},\mathbf{\Omega}\succeq\mathbf{0}\label{per7:1e}.
\end{align}
\end{subequations}
Note that the rank constraint in $(\ref{per7:1e})$ is non-convex. By dropping the rank-one constraint in $(\ref{per7:1e})$, the SDR of problem $(\ref{per7:1})$ can be expressed as
\begin{align}\label{per8}
\phi(\beta)=&\max_{{\tilde{\mathbf{W}},\mathbf{\Omega}}}~ \frac{P_s\mathrm{Tr}(\mathbf{A}_1\tilde{\mathbf{W}})}{\sigma^2\mathrm{Tr}(\mathbf{A}_2\tilde{\mathbf{W}})+\mathbf{h}^H(\theta_{rd})\mathbf{\Omega}\mathbf{h}(\theta_{rd})+\sigma^2}\nonumber\\
&\mathrm{s.t.}~(\ref{per7:1b}),(\ref{per7:1c}),(\ref{per7:1d}),\tilde{\mathbf{W}}\succeq\mathbf{0},\mathbf{\Omega}\succeq\mathbf{0}.
\end{align}
It can be observed that $(\ref{per8})$ constitutes a quasi-convex problem, which can be transformed into a convex optimization problem by using the Charnes-Cooper transformation \cite{Charnes}. Upon introducing slack variable $\tau$, problem $(\ref{per8})$ can be equivalently rewritten as
\begin{align}\label{per9}
&\phi(\beta)=\max_{{\mathbf{Q},\mathbf{\Upsilon}},\tau}~ P_s\mathrm{Tr}(\mathbf{A}_1\mathbf{Q})\nonumber\\
&\mathrm{s.t.}~P_s\mathrm{Tr}(\mathbf{B}_{1_m}\mathbf{Q})-\beta\sigma^2\mathrm{Tr}(\mathbf{B}_{2_m}\mathbf{Q})-\beta\mathrm{Tr}(\bar{\mathbf{H}}_{re_m}\mathbf{\Upsilon})\nonumber\\
&~~~~~-\beta\sigma^2\tau\leq 0\nonumber,m\in \mathcal{M},\\
&~~~~~P_s\mathrm{Tr}(\mathbf{C}_1\mathbf{Q})+\sigma^2\mathrm{Tr}(\mathbf{Q})+\mathrm{Tr}(\mathbf{\Upsilon})\leq P_t\tau, \nonumber\\
&~~~~~\sigma^2\mathrm{Tr}(\mathbf{A}_2\mathbf{Q})+\mathbf{h}^H(\theta_{rd})\mathbf{\Upsilon}\mathbf{h}(\theta_{rd})+\sigma^2\tau=1,\nonumber\\
&~~~~~P_s\mathrm{Tr}(\mathbf{D}_1\mathbf{Q})+\sigma^2\mathrm{Tr}(\mathbf{D}_2\mathbf{Q})+\mathbf{h}^H(\theta_{rp})\mathbf{\Upsilon}\mathbf{h}(\theta_{rp})\nonumber\\
&~~~~~\geq \frac{P_{min}\tau}{\rho},\mathbf{Q}\succeq\mathbf{0},\mathbf{\Upsilon}\succeq\mathbf{0},\tau>0,
\end{align}
where $\mathbf{Q}=\tilde{\mathbf{W}}\tau$ and $\mathbf{\Upsilon}=\mathbf{\Omega}\tau$. Since problem $(\ref{per9})$ is a standard SDP problem \cite{Boyd}, its optimal solution can be found by using SDP solvers, such as CVX. If the optimal solution of problem $(\ref{per9})$ is $(\mathbf{Q}^\star,\mathbf{\Upsilon}^\star, \tau)$, then $(\mathbf{Q}^\star/\tau,\mathbf{\Upsilon}^\star/\tau)$ will be the optimal solution of problem $(\ref{per8})$.

Since we have dropped the rank-one constraint in the problem $(\ref{per7:1})$ and reformulated it as a SDR problem $(\ref{per8})$, the optimal solution of $(\ref{per8})$ may not be rank-one and thus the optimal objective value of $(\ref{per8})$ generally serves an upper bound of $(\ref{per7:1})$. Next, we show that the above SDR is in fact tight. We consider the power minimization problem as follows
\begin{align}\label{per10}
&\min_{{\tilde{\mathbf{W}},\mathbf{\Omega}}}~ P_s\mathrm{Tr}(\mathbf{C}_1\tilde{\mathbf{W}})+\sigma^2\mathrm{Tr}(\tilde{\mathbf{W}})\nonumber\\
&\mathrm{s.t.}\frac{P_s\mathrm{Tr}(\mathbf{A}_1\tilde{\mathbf{W}})}{\sigma^2\mathrm{Tr}(\mathbf{A}_2\tilde{\mathbf{W}})+\mathbf{h}^H(\theta_{rd})\mathbf{\Omega}\mathbf{h}(\theta_{rd})+\sigma^2}\geq \phi(\beta),\nonumber\\
&~~~~~(\ref{per7:1b}),(\ref{per7:1c}),(\ref{per7:1d}),\tilde{\mathbf{W}}\succeq\mathbf{0},\mathbf{\Omega}\succeq\mathbf{0},
\end{align}
where $\phi(\beta)$ is the optimal value of problem $(\ref{per8})$. Observe that the optimal solution of problem $(\ref{per10})$ is also an optimal solution of $(\ref{per8})$. The proof is similar to that in \cite{Li2015} and thus omitted here for brevity.
In order to obtain the optimal solution of $(\ref{per7:1})$, we should first obtain the optimal solution $(\tilde{\mathbf{W}}^\star, \mathbf{\Omega}^\star)$ and the optimal value $\phi(\beta)$ of problem $(\ref{per8})$ by solving $(\ref{per9})$. If $\mathrm{rank}(\tilde{\mathbf{W}}^\star)=1$, then we get the optimal solution of $(\ref{per7:1})$. Otherwise, the rank-one solution can be found by solving $(\ref{per10})$.

\emph{Lemma 1:} The optimal solution $\tilde{\mathbf{W}}^\star$ in $(\ref{per10})$  satisfies $\mathrm{rank}(\tilde{\mathbf{W}}^\star)=1$.

\emph{Proof}: See Appendix B.

Since $\tilde{\mathbf{W}}^\star$ is a rank-one matrix, we can write $\tilde{\mathbf{W}}^\star=\mathbf{w}^\star\mathbf{w}^{\star H}$ by using eigenvalue decomposition. Thus, the SDR is tight and the optimal solution of $(\ref{per3})$ is $\mathbf{w}^\star$ and $\mathbf{\Omega}^\star$. Up to now, we have solved the problem $(\ref{per2})$.

Let us now return to the procedure used for the problem $(\ref{per1})$. The maximum of $\beta$ should be found by a 1D search. According to the fact that the secrecy rate is always higher than or equal to zero, we get
\begin{align}\label{per5}
\beta&\leq\frac{P_s\mathbf{w}^H\mathbf{A}_1\mathbf{w}}{\sigma^2\mathbf{w}^H\mathbf{A}_2\mathbf{w}+\mathbf{h}^H(\theta_{rd})\mathbf{\Omega}\mathbf{h}(\theta_{rd})+\sigma^2}\nonumber\\
&\leq\frac{P_s\mathbf{w}^H\mathbf{A}_1\mathbf{w}}{\sigma^2\mathbf{w}^H\mathbf{A}_2\mathbf{w}+\sigma^2}.
\end{align}
From the transmit power constraint in $(\ref{per3})$, we have $\frac{\sigma^2}{P_t}\mathbf{w}^H\mathbf{w}\leq 1$, hence
\begin{align}\label{per6}
\beta\leq\frac{P_s\mathbf{w}^H\mathbf{A}_1\mathbf{w}}{\sigma^2\mathbf{w}^H\mathbf{A}_2\mathbf{w}+\frac{\sigma^4}{P_t}\mathbf{w}^H\mathbf{w}}=\frac{P_s\mathbf{w}^H\mathbf{A}_1\mathbf{w}}{\mathbf{w}^H(\sigma^2\mathbf{A}_2+\frac{\sigma^4}{P_t}\mathbf{I}_{N^2})\mathbf{w}}.
\end{align}
Observe that $\mathbf{A}_1$ can be recast as
\begin{align}\label{A1}
&\mathbf{A}_1=\left[\mathbf{h}^*(\theta_{sr})\otimes\mathbf{h}(\theta_{rd})\right] \left[\mathbf{h}^T(\theta_{sr})\otimes\mathbf{h}^H(\theta_{rd})\right]=\mathbf{a}_1\mathbf{a}_1^H,
\end{align}
where $\mathbf{a}_1\in\mathbb{C}^{N^2\times1}$. Therefore, we have $\mathrm{rank}(\mathbf{A}_1)=1$. According to $(\ref{per6})$ and $(\ref{A1})$, the upper bound of $\beta$ is given by
\begin{align}\label{betamax}
\beta\leq P_s\mathbf{a}_1^H(\sigma^2\mathbf{A}_2+\frac{\sigma^4}{P_t}\mathbf{I}_{N^2})^{-1}\mathbf{a}_1=\beta_{max}.
\end{align}
The proposed SRM-1D scheme is summarized in Algorithm 1.

\begin{algorithm}
\begin{algorithmic}
\STATE Initialize $\varepsilon$, $n$, $\beta$, and compute $\beta_{max}$.
\REPEAT
\STATE 1) Set $n=n+1$, $\beta=\beta+\varepsilon$.
\STATE 2) Solve problem $(\ref{per9})$ and obtain the optimal solution
$~~~~$$(\tilde{\mathbf{Q}}^\star(n), \mathbf{\Upsilon}^\star(n), \tau^\star(n))$ and optimal value $\phi(\beta)(n)$.
\STATE 3) Compute secrecy rate $R_s(n)$ according to the objective $~~~~$function of $(\ref{per1})$.
\UNTIL $\beta>\beta_{max}$.
\STATE $\bullet$  $n=\mathrm{arg} \max\limits_{n} Rs(n)$, and $\tilde{\mathbf{W}}^\star=\mathbf{Q}^\star(n)/\tau^\star(n)$,  \\ $~~$$\mathbf{\Omega}^\star=\tilde{\mathbf{\Upsilon}}^\star(n)/\tau^\star(n)$. If rank($\tilde{\mathbf{W}}^\star$)=1, then
go to next step; $~~~$otherwise, solve $(\ref{per10})$.
\STATE $\bullet$ By using eigenvalue decomposition, we can obtain $\mathbf{w}^\star$,
and reconstruct $\mathbf{W}^\star$;  $\mathbf{z}={\mathbf{\Omega}^\star}^{\frac{1}{2}}\mathbf{v}$ and $\mathbf{v}\sim\mathcal {C}\mathcal {N}(0, \mathbf{I}_N)$.
\RETURN $\mathbf{W}^\star$ and $\mathbf{z}^\star$.
\end{algorithmic}
\caption{Maximize secrecy rate based on 1D search}\label{algorithm 1}
\end{algorithm}
\subsection{Maximization of Signal-to-Leakage-AN-Noise-Ratio (Max-SLANR) Scheme}\label{sub_slanr}
\newcounter{mytempeqncnt1}
\begin{figure*}[!t]
\normalsize
\setcounter{mytempeqncnt1}{\value{equation}}
\begin{align}\label{SL2}
&\max_{{\mathbf{W},\mathbf{\Omega}}}~ \frac{P_s|\mathbf{h}^H(\theta_{rd})\mathbf{W}\mathbf{h}(\theta_{sr})|^2}{\sum_{m=1}^M\mathbb{E}[P_s|\mathbf{h}^H(\theta_{re_m})\mathbf{W}\mathbf{h}(\theta_{sr})|^2]+\mathbf{h}^H(\theta_{rd})\mathbf{\Omega}\mathbf{h}(\theta_{rd})+\sigma^2\|\mathbf{W}^H\mathbf{h}(\theta_{rd})\|^2+\sigma^2}\nonumber\\
&\mathrm{s.t.}~P_s\|\mathbf{W}\mathbf{h}(\theta_{sr})\|^2+\sigma^2\mathrm{Tr}(\mathbf{W}^H\mathbf{W})+\mathrm{Tr}(\mathbf{\Omega})\leq P_t, \nonumber\\
&~~~~~P_s|\mathbf{h}^H(\theta_{rp})\mathbf{W}\mathbf{h}(\theta_{sr})|^2+\sigma^2\|\mathbf{W}^H\mathbf{h}(\theta_{rp})\|^2+\mathbf{h}^H(\theta_{rp})\mathbf{\Omega}\mathbf{h}(\theta_{rp})\geq \frac{P_{min}}{\rho},\mathbf{\Omega}\succeq\mathbf{0}.
\tag{33}
\end{align}
\begin{align}\label{SL3}
&\max_{{\tilde{\mathbf{W}},\mathbf{\Omega}}}~ \frac{P_s\mathrm{Tr}(\mathbf{A}_1\tilde{\mathbf{W}})}{\sum_{m=1}^MP_s\mathrm{Tr}(\mathbf{B}_{1_m}\tilde{\mathbf{W}})+\mathbf{h}^H(\theta_{rd})\mathbf{\Omega}\mathbf{h}(\theta_{rd})+\sigma^2\mathrm{Tr}(\mathbf{A}_2\tilde{\mathbf{W}})+\sigma^2}\nonumber\\
&\mathrm{s.t.}~P_s\mathrm{Tr}(\mathbf{C}_1\tilde{\mathbf{W}})+\sigma^2\mathrm{Tr}(\tilde{\mathbf{W}})+\mathrm{Tr}(\mathbf{\Omega})\leq P_t, \nonumber\\
&~~~~~P_s\mathrm{Tr}(\mathbf{D}_1\tilde{\mathbf{W}})+\sigma^2\mathrm{Tr}(\mathbf{D}_2\tilde{\mathbf{W}})+\mathbf{h}^H(\theta_{rp})\mathbf{\Omega}\mathbf{h}(\theta_{rp})\geq \frac{P_{min}}{\rho},\tilde{\mathbf{W}}\succeq\mathbf{0}, \mathbf{\Omega}\succeq\mathbf{0}.
\tag{34}
\end{align}
\setcounter{equation}{\value{mytempeqncnt1}}
\hrulefill
\vspace*{4pt}
\end{figure*}
In the previous subsection, we employed a 1D search and a SDR-based algorithm to solve problem (P1). Although we have already derived $\beta_{max}$, to limit the range of the 1D search, the complexity of the 1D search still remains high since for each $\beta$, a SDP with $\mathcal{O}(N^{13})$ needs to be solved. In order to avoid employing the 1D search, we propose an alternative algorithm for the suboptimal solution of (P1). Specifically, we propose an algorithm to maximize the SLANR rather than secrecy rate, subject to the total power and to the harvested energy constraints. Based on the concept of leakage \cite{Shu2016Adaptive}, from $(\ref{y0})$ and $(\ref{ye})$, the optimization problem (P1) can be reformulated as (\ref{SL2}) at the top of the next page. The numerator of the objective function in $(\ref{SL2})$ represents the received confidential message power at the IR, and the first term in the denominator denotes the sum of confidential message power leaked to all eavesdroppers.

Following similar steps as in Section \ref{sub_SRM} and dropping the rank-one constraint, the related SDR problem can be formulated as show in $(\ref{SL3})$ at the top of the page, where $\tilde{\mathbf{W}}=\mathbf{w}\mathbf{w}^H\in\mathbb{C}^{N^2\times N^2}$ and $\mathbf{w}= \mathrm{vec}(\mathbf{W})\in \mathbb{C}^{N^2\times 1}$.
Note that all constraints in $(\ref{SL3})$ are convex. However, the objective function is a linear fractional function, which is quasi-convex. Similar to $(\ref{per8})$, we transform $(\ref{SL3})$  into a convex optimization problem  by using the Charnes-Cooper transformation\cite{Charnes}. Problem $(\ref{SL3})$ can then be equivalently rewritten as
\setcounter{equation}{34}
\begin{align}\label{SL4}
&\max_{{\mathbf{Q},\mathbf{\Upsilon}},\tau}~ P_s\mathrm{Tr}(\mathbf{A}_1\mathbf{Q})\nonumber\\
&\mathrm{s.t.}~\sum_{m=1}^MP_s\mathrm{Tr}(\mathbf{B}_{1_m}\mathbf{Q})+\mathbf{h}^H(\theta_{rd})\mathbf{\Upsilon}\mathbf{h}(\theta_{rd})+\nonumber\\
&~~~~~\sigma^2\mathrm{Tr}(\mathbf{A}_2\mathbf{Q})+\sigma^2\tau=1,\nonumber\\
&~~~~~P_s\mathrm{Tr}(\mathbf{C}_1\mathbf{Q})+\sigma^2\mathrm{Tr}(\mathbf{Q})+\mathrm{Tr}(\mathbf{\Upsilon})\leq P_t\tau, \nonumber\\
&~~~~~P_s\mathrm{Tr}(\mathbf{D}_1\mathbf{Q})+\sigma^2\mathrm{Tr}(\mathbf{D}_2\mathbf{Q})+\mathbf{h}^H(\theta_{rp})\mathbf{\Upsilon}\mathbf{h}(\theta_{rp})\nonumber\\
&~~~~~\geq \frac{P_{min}\tau}{\rho},\mathbf{Q}\succeq\mathbf{0},\mathbf{\Upsilon}\succeq\mathbf{0},\tau>0,
\end{align}
where $\tau$ is a slack variable, $\mathbf{Q}=\tilde{\mathbf{W}}\tau$ and $\mathbf{\Upsilon}=\mathbf{\Omega}\tau$. To prove that the relaxation is tight, we consider the associated power minimization problem, which is similar to that in Section \ref{sub_SRM}, yielding
\begin{align}\label{SL5}
&\min_{{\tilde{\mathbf{W}},\mathbf{\Omega}}}~
P_s\mathrm{Tr}(\mathbf{C}_1\tilde{\mathbf{W}})+\sigma^2\mathrm{Tr}(\tilde{\mathbf{W}})\nonumber\\
&\mathrm{s.t.}~-P_s\mathrm{Tr}(\mathbf{A}_1\tilde{\mathbf{W}})+\phi\sum_{m=1}^MP_s\mathrm{Tr}(\mathbf{B}_{1_m}\tilde{\mathbf{W}})+\nonumber\\
&~~~~~\phi\mathbf{h}^H(\theta_{rd})\mathbf{\Omega}\mathbf{h}(\theta_{rd})+\phi\sigma^2\mathrm{Tr}(\mathbf{A}_2\tilde{\mathbf{W}})+\phi\sigma^2\leq 0,\nonumber\\
&~~~~~P_s\mathrm{Tr}(\mathbf{C}_1\tilde{\mathbf{W}})+\sigma^2\mathrm{Tr}(\tilde{\mathbf{W}})+\mathrm{Tr}(\mathbf{\Omega})\leq P_t, \nonumber\\
&~~~~~P_s\mathrm{Tr}(\mathbf{D}_1\tilde{\mathbf{W}})+\sigma^2\mathrm{Tr}(\mathbf{D}_2\tilde{\mathbf{W}})+\mathbf{h}^H(\theta_{rp})\mathbf{\Omega}\mathbf{h}(\theta_{rp})\nonumber\\
&~~~~~\geq \frac{P_{min}}{\rho},\mathbf{\Omega}\succeq\mathbf{0},\tilde{\mathbf{W}}\succeq\mathbf{0},
\end{align}
where $\phi$ is the optimal value of $(\ref{SL4})$. Problem $(\ref{SL5})$ is a standard SDP problem.

\emph{Lemma 2:} The optimal solution $\tilde{\mathbf{W}}^\star$ in $(\ref{SL5})$ satisfies $\mathrm{rank}(\tilde{\mathbf{W}}^\star)=1$.

\emph{Proof}: See Appendix C.

\subsection{Low-complexity SCA Scheme}
In the \ref{sub_SRM} and \ref{sub_slanr}, we have proposed the SRM-1D and the Max-SLANR schemes to obtain the information beamforming matrix and the AN covariance matrix. Both of the two schemes have high computational complexity because their optimization variables are matrices. To facilitate implementation in practice, we propose a low complexity scheme based on SCA in this subsection. Specifically, we first formulate the optimization problem, then convert it into the SOCP problem, and use the SCA method to solve the problem iteratively. Different from designing the AN covariance matrix $\mathbf{\Omega}$ in the previous two subsections, here we are devoted to designing the AN beamforming vector $\mathbf{v}$, where $\mathbf{\Omega}=\mathbf{v}\mathbf{v}^H$.

The optimization problem (\ref{P1:1}) can be rewritten as
\begin{subequations}\label{SCA1}
\begin{align}
&\max_{{\mathbf{w},\mathbf{v}}}\min_m~\frac{1+\mathrm{SINR}_d}{1+\mathbb{E}[\mathrm{SINR}_{e_m}]}\label{SCA1a}\\ &\mathrm{s.t.}~\mathbf{w}^H(P_s\mathbf{C}_1+\sigma^2\mathbf{I}_{N^2})\mathbf{w}+\mathbf{v}^H\mathbf{v}\leq P_t, \label{SCA1b}\\
&~~~~~\mathbf{w}^H(P_s\mathbf{D}_1+\sigma^2\mathbf{D}_2)\mathbf{w}+\mathbf{v}^H\mathbf{h}(\theta_{rp})\mathbf{h}^H(\theta_{rp})\mathbf{v}\geq\frac{P_{min}}{\rho},\label{SCA1c} \end{align}
\end{subequations}
where $\mathrm{SINR}_d$ and $\mathbb{E}[\mathrm{SINR}_{e_m}]$ are defined in (\ref{per3a}) and (\ref{per3b}), respectively.
By introducing slack variables $r_1$ and $r_2$, problem (\ref{SCA1}) is equivalently rewritten as
\begin{subequations}\label{SCA2}
\begin{align}
&\max_{{\mathbf{w},\mathbf{v},r_1,r_2}}~r_1r_2\\
&\mathrm{s.t.}~1+\mathrm{SINR}_d\geq r_1,\label{SCA2b}\\
&~~~~~1+\mathbb{E}[\mathrm{SINR}_{e_m}]\leq \frac{1}{r_2},~m\in\mathcal{M},\label{SCA2c}\\
&~~~~~(\ref{SCA1b}),(\ref{SCA1c}).
\end{align}
\end{subequations}
(\ref{SCA2b}) and (\ref{SCA2c}) can be rearranged as
\begin{subequations}\label{SCA3}
\begin{align}
&\sigma^2\mathbf{w}^H\mathbf{A}_2\mathbf{w}+\mathbf{v}^H\mathbf{h}(\theta_{rd})\mathbf{h}^H(\theta_{rd})\mathbf{v}+\sigma^2\leq\frac{P_s\mathbf{w}^H\mathbf{A}_1\mathbf{w}}{r_1-1},\label{SCA3a}\\
&\mathbf{w}^H(P_s\mathbf{B}_{1_m}+\sigma^2\mathbf{B}_{2_m})\mathbf{w}+\mathbf{v}^H\mathbf{H}_{re_m}\mathbf{v}+\sigma^2 \nonumber\\ &~\leq\frac{1}{{r_2}}(\sigma^2\mathbf{w}^H\mathbf{B}_{2_m}\mathbf{w}+\mathbf{v}^H\mathbf{H}_{re_m}\mathbf{v}+\sigma^2),m\in\mathcal{M}\label{SCA3b},
\end{align}
\end{subequations}
respectively. Since the quadratic-over-linear function is convex\cite{Boyd}, the right-hand-side (RHS) of (\ref{SCA3a}) and (\ref{SCA3b}) are convex functions ($r_1>1,r_2>0$). In the following, we first transform the (\ref{SCA3a}) and (\ref{SCA3b}) into convex constraints by using the first-order Talyor expansions \cite{SecreyrateVPoor}, and then convert them into the second-order cone (SOC) constraints. To this end, we define
\begin{align}\label{SCA4}
f_{\mathbf{A},a}(\mathbf{x},r)=\frac{\mathbf{x}^H\mathbf{A}\mathbf{x}}{r-a},
\end{align}
where $\mathbf{A}\succeq\mathbf{0}$ and $r>a$. We perform a first-order Taylor expansion on (\ref{SCA4}) at point $(\mathbf{\tilde{x}},\tilde{r})$\cite{Li2014Secure}, yielding:
\begin{align}\label{SCA5}
f_{\mathbf{A},a}(\mathbf{x},r)\geq& F_{\mathbf{A},a}(\mathbf{x},r,\mathbf{\tilde{x}},\tilde{r})\nonumber\\
=&\frac{2\mathrm{Re}\{\mathbf{\tilde{x}}^H\mathbf{A}\mathbf{x}\}}{\tilde{r}-a}-\frac{\mathbf{\tilde{x}}^H\mathbf{A}\mathbf{\tilde{x}}}{(\tilde{r}-a)^2}(r-a),
\end{align}
where the inequality holds due to the convexity of $f_{\mathbf{A},a}(\mathbf{x},r)$ with respect to $\mathbf{x}$ and $r$. Therefore, (\ref{SCA3a}) and (\ref{SCA3b}) can be rewritten as
\begin{subequations}\label{SCA6}
\begin{align}
&\sigma^2\mathbf{w}^H\mathbf{A}_2\mathbf{w}+\mathbf{v}^H\mathbf{h}(\theta_{rd})\mathbf{h}^H(\theta_{rd})\mathbf{v}+\sigma^2\nonumber\\
&\leq P_sF_{\mathbf{A}_1,1}(\mathbf{w},r_1,\mathbf{\tilde{w}},\tilde{r}_1),\label{SCA6a}\\
&\mathbf{w}^H(P_s\mathbf{B}_{1_m}+\sigma^2\mathbf{B}_{2_m})\mathbf{w}+\mathbf{v}^H\mathbf{H}_{re_m}\mathbf{v}+\sigma^2\leq \mathfrak{F}_m,\nonumber\\
&~m\in\mathcal{M}\label{SCA6b},
\end{align}
\end{subequations}
which can be transformed into the SOC constraints, i.e.,
\begin{subequations}\label{SCA7}
\begin{align}
&\left\|\big[2\sigma\mathbf{A}_2^{\frac{1}{2}}\mathbf{w};2\mathbf{h}^H(\theta_{rd})\mathbf{v};2\sigma;P_sF_{\mathbf{A}_1,1}(\mathbf{w},r_1,\mathbf{\tilde{w}},\tilde{r}_1)-1\big]\right\|\nonumber\\
&\leq P_sF_{\mathbf{A}_1,1}(\mathbf{w},r_1,\mathbf{\tilde{w}},\tilde{r}_1)+1,\label{SCA7a}\\
&\left\|\big[2(P_s\mathbf{B}_{1_m}+\sigma^2\mathbf{B}_{2_m})^{\frac{1}{2}}\mathbf{w};2\mathbf{H}_{re_m}^{\frac{1}{2}}\mathbf{v};2\sigma;\mathfrak{F}_m-1\big]\right\|\nonumber\\
&\leq \mathfrak{F}_m+1,~m\in\mathcal{M},\label{SCA7b}
\end{align}
\end{subequations}
where $\mathfrak{F}_m$ is defined as
\begin{align}\label{SCA8}
\mathfrak{F}_m=&\sigma^2F_{\mathbf{B}_{2_m},0}(\mathbf{w},r_2,\mathbf{\tilde{w}},\tilde{r}_2)+F_{\mathbf{H}_{re_m},0}(\mathbf{v},r_2,\mathbf{\tilde{v}},\tilde{r}_2)\nonumber\\
&+\sigma^2\left(\frac{2}{\tilde{r}_2}-\frac{r_2}{\tilde{r}_2^2}\right).
\end{align}
It is easy to see that the objective function of the problem (\ref{SCA2}) is non-concave and the constraint (\ref{SCA1c}) is non-convex. To handle the non-concave objective function, we introduce slack variables $t$ and $\psi$ and then rewrite the problem (\ref{SCA2}) as
\begin{subequations}\label{SCA9}
\begin{align}
&\max_{{\mathbf{w},\mathbf{v},r_1,r_2,t,\psi}}~t \label{SCA9a}\\
&\mathrm{s.t.}~r_1r_2\geq \psi^2,~\psi^2\geq t,\label{SCA9b}.\\
&~~~~~(\ref{SCA7a}),(\ref{SCA7b}),(\ref{SCA1b}),(\ref{SCA1c})\label{SCA9c}.
\end{align}
\end{subequations}
Note that the first term of the (\ref{SCA9b}) can be rearranged as the SOC constraint, i.e.,
\begin{align}\label{SCA10}
\|[r_1-r_2;2\psi]\|\leq r_1+r_2.
\end{align}
For the second term of the (\ref{SCA9b}), we employ the first-order Taylor expansion at the point $\tilde{\psi}$ and transform it into the linear constraint, i.e.,
\begin{align}\label{SCA11}
2\tilde{\psi}\psi-\tilde{\psi}^2\geq t.
\end{align}

In the following, we will focus on dealing with the non-convex constraint (\ref{SCA1c}). To convert (\ref{SCA1c}) into the convex constraint, we define
\begin{align}\label{SCA12}
u_{\mathbf{A}}(\mathbf{x})=\mathbf{x}^H\mathbf{A}\mathbf{x},
\end{align}
where $\mathbf{A}\succeq\mathbf{0}$. Since $u_{\mathbf{A}}(\mathbf{x})$ is a convex function, we have the following inequality
\begin{align}\label{SCA13}
u_{\mathbf{A}}(\mathbf{x})\geq U_{\mathbf{A}}(\mathbf{x},\mathbf{\tilde{x}})=2\mathrm{Re}(\mathbf{\tilde{x}}^H\mathbf{A}\mathbf{x})-\mathbf{\tilde{x}}^H\mathbf{A}\mathbf{\tilde{x}},
\end{align}
where the inequality (\ref{SCA13}) holds based on the first-order Taylor expansion at the point $\mathbf{\tilde{x}}$. According to (\ref{SCA13}), (\ref{SCA1c}) can be rewritten as
\begin{align}\label{SCA14}
U_{\mathbf{G}}(\mathbf{w},\mathbf{\tilde{w}})+U_{\mathbf{H}_{rp}}(\mathbf{v},\mathbf{\tilde{v}})\geq \frac{P_{min}}{\rho},
\end{align}
where $\mathbf{G}=P_s\mathbf{D}_1+\sigma^2\mathbf{D}_2$ and $\mathbf{H}_{rp}=\mathbf{h}(\theta_{rp})\mathbf{h}^H(\theta_{rp})$. In addition, (\ref{SCA1b}) can be equivalently rewritten as
\begin{align}\label{SCA15}
\left\|\big[(P_s\mathbf{C}_1+\sigma^2\mathbf{I}_{N^2})^{\frac{1}{2}}\mathbf{w};\mathbf{v}\big]\right\|\leq \sqrt{P_t}.
\end{align}
According to the above transformation of the objective function and the constraints of problem (\ref{SCA1}), we can convert (\ref{SCA1}) into the following SOCP problem
\begin{align}\label{SCA16}
&\max_{{\mathbf{w},\mathbf{v},r_1,r_2,t,\psi}}~t \nonumber\\
&\mathrm{s.t.}~(\ref{SCA10}),(\ref{SCA11}),(\ref{SCA14}),(\ref{SCA15}),(\ref{SCA7a}),(\ref{SCA7b}).
\end{align}
It can be seen that the optimization problem (\ref{SCA16}) consists of a linear objective function and several SOC constraints. Therefore, problem (\ref{SCA16}) is a convex optimization problem. For a given feasible solution ($\mathbf{\tilde{w}},\mathbf{\tilde{v}},\tilde{r}_1,\tilde{r}_2,\tilde{\psi}$), we can solve the problem (\ref{SCA16}) by means of  convex optimization tools such as CVX \cite{Boyd}. Based on the idea of SCA, the original optimization problem (\ref{SCA1}) can be solved iteratively by solving a series of convex subproblems (\ref{SCA16}).
The current optimal solution of the convex subproblem (\ref{SCA16}) is gradually approaching the optimal solution of the original problem with the increase of the number of iterations, until the algorithm converges\cite{Zappone2017Globally}. Algorithm $2$ lists the detailed process of the SCA algorithm.
\begin{algorithm}
\begin{algorithmic}
\STATE Initialize: Given a feasible solution ($\mathbf{\tilde{w}}^0,\mathbf{\tilde{v}}^0,\tilde{r}_1^0,\tilde{r}_2^0,\tilde{\psi}^0$); n=0.
\REPEAT
\STATE 1. Solve the problem (\ref{SCA16}) with ($\mathbf{\tilde{w}}^n,\mathbf{\tilde{v}}^n,\tilde{r}_1^n,\tilde{r}_2^n,\tilde{\psi}^n$) and obtain the current optimal solution ($\mathbf{\tilde{w}}^*,\mathbf{\tilde{v}}^*,\tilde{r}_1^*,\tilde{r}_2^*,\tilde{\psi}^*$); n=n+1.
\STATE 2. Update ($\mathbf{\tilde{w}}^n,\mathbf{\tilde{v}}^n,\tilde{r}_1^n,\tilde{r}_2^n,\tilde{\psi}^n$)=($\mathbf{\tilde{w}}^*,\mathbf{\tilde{v}}^*,\tilde{r}_1^*,\tilde{r}_2^*,\tilde{\psi}^*$).
\STATE 3. Compute secrecy rate $\bar{R}_s^n$.
\UNTIL $|\bar{R}_s^{n}-\bar{R}_s^{n-1}|<\delta$ is met, where $\delta$ denotes the convergence tolerance.
\end{algorithmic}
\caption{SCA Algorithm for Solving Problem $(\ref{SCA1})$}\label{algorithm 2}
\end{algorithm}

\subsection{Complexity Analysis}
\begin{table*}
\centering
  \caption{Complexity analysis of proposed algorithms}
\begin{tabular}{|c|c|}
\hline  
Algorithms & Complexity order \big(suppressing $\ln(\frac{1}{\epsilon})$\big)\\
\hline  
SRM-1D-Robust &$\mathcal{O}\Big(nT\sqrt{N^2+N+M+5}\big(N^6+N^3+n(N^4+N^2+M+5)+M+5+n^2\big)\Big)$, where $n=\mathcal{O}(N^4+N^2+1)$.\\
\hline 
Max-SLANR-Robust&$\mathcal{O}\Big(n\sqrt{N^2+N+5}\big(N^6+N^3+n(N^4+N^2+5)+5+n^2\big)\Big)$, where $n=\mathcal{O}(N^4+N^2+1)$.\\
\hline 
SCA-Robust&$\mathcal{O}\Big(nL\sqrt{2M+8}\big((N^2+N)^2+(M+1)(N^2+N+1)^2+6+2n+n^2\big)\Big)$, where $n=\mathcal{O}(N^2+N+4)$.\\
\hline 
\end{tabular}
\end{table*}
In this section, we analyze and compare the complexity of the proposed three schemes in the previous three subsections. For the SRM-1D scheme, we convert the SRM problem to the SDP form and solve it with an 1D search. The complexity of each search is calculated according to problem (\ref{per9}). Problem (\ref{per9}) consists of $M+5$ linear constraints with dimension $1$, one LMI constraint of size $N^2$, and one LMI constraint of size $N$. The number of decision variables $n$ is on the order of $N^4+N^2+1$. Therefore, the total complexity based on the SRM-1D scheme can be expressed as \cite{KY2014Outage}
\begin{align}\label{COMP1}
&\mathcal{O}\Big(nT\sqrt{N^2+N+M+5}\big(N^6+N^3+n(N^4+N^2+\nonumber\\
&M+5)+M+5+n^2\big)\ln\left(1/\epsilon\right)\Big),
\end{align}
where $T$ denotes the number of iterations in the 1D search and $\epsilon$ denotes the computation accuracy.

For the Max-SLANR scheme, we compute the complexity of the optimization problem (\ref{SL4}),
which consists of one LMI constraint with size of $N^2$, one LMI constraint with size of $N$ and five linear constraints. The number of decision variables $n$ is on the order of $N^4+N^2+1$. Therefore, the total complexity based on the Max-SLANR scheme can be expressed as
\begin{align}\label{COMP2}
&\mathcal{O}\Big(n\sqrt{N^2+N+5}\big(N^6+N^3+n(N^4+N^2+5)\nonumber\\
&+5+n^2\big)\ln\left(1/\epsilon\right)\Big),
\end{align}

For the proposed SCA scheme, we first formulate the SRM problem, then convert it into the SOCP form, and use the SCA algorithm to solve it iteratively. The complexity of each iteration is calculated according to problem (\ref{SCA16}). Problem (\ref{SCA16}) includes two linear constraints, one SOC constraint of dimension $2$, one SOC constraint of dimension $N^2+N$, M+1 SOC constraints of dimension $N^2+N+1$. The number of decision variables $n$ is on the order of $N^2+N+4$. Therefore, the total complexity based on the SCA scheme is given by
\begin{align}\label{COMP3}
&\mathcal{O}\Big(nL\sqrt{2M+8}\big((N^2+N)^2+(M+1)(N^2+N+1)^2\nonumber\\
&+6+2n+n^2\big)\ln\left(1/\epsilon\right)\Big),
\end{align}
where $L$ is the number of iterations. The complexity of the proposed algorithms are also listed in Table I at the top the page.

\emph{Discussions:} Upon comparing $(\ref{COMP1})$, $(\ref{COMP2})$ and $(\ref{COMP3})$, it can be observed that the complexity of the SCA algorithm is much lower than that of the SRM-1D and the Max-SLANR schemes. The complexity of Max-SLANR scheme is slightly less than that of the SRM-1D scheme, but the SLANR scheme does not require 1D search. Moreover, the complexity of the SRM-1D scheme grows linearly upon increasing the precision of the 1D search, with the number of iterations $T$, and it grows with the number $M$ of eavesdroppers, while the complexity of the Max-SLANR scheme is not related to either of them. This implies that when the number of eavesdroppers increases, the complexity of the Max-SLANR scheme remains constant, while the complexity of the SRM-1D and SCA schemes increases. For example, for a system with $N=6$, $M=2$, $T=12$ and $L=6$, the complexity of the SRM-1D, the Max-SLANR, and SCA schemes, are $\mathcal{O}(3.98\times 10^{11})$, $\mathcal{O}(3.25\times 10^{10})$, and $\mathcal{O}(8.87\times 10^{6})$, respectively. Therefore, the complexity of the SCA scheme is much lower than that of the other two schemes.

\section{Simulation Results}\label{sec_simu}
\begin{table}[!ht]
\centering
  \caption{Simulation parameter}
\begin{tabular}{|c|c|}
\hline  
Parameters & Values\\
\hline  
The transmit power at the relay ($P_t$)&$30$dBm\\
\hline 
The transmit power at the source ($P_s$)&$30$dBm\\
\hline 
the noise variance ($\sigma^2$) &$-10$dBm\\
\hline 
The number of transmit antennas at the relay ($N$)&$6$\\
\hline 
The number of eavesdroppers ($M$)&$2$\\
\hline 
the minimum energy required by the ER ($P_{min}$)&$10$dBm\\
\hline 
The maximum angle estimation error ($\triangle\theta_{\mathrm{max}}$) &$6^\circ$\\
\hline 
The normalization factor ($K_e$) &$0.9$\\
\hline 
The energy transfer efficiency ($\rho$) &$0.8$\\
\hline 
 The direction angle of the source ($\theta_{sr}$) &$-\frac{7\pi}{18}$\\
\hline 
The direction angle of the IR ($\theta_{rd}$) &$\frac{\pi}{2}$\\
\hline 
The direction angle of the eavesdroppers ($\theta_{re_1},\theta_{re_2}$) &$\frac{\pi}{3},\frac{11\pi}{18}$\\
\hline 
The convergence tolerance ($\delta$) & $10^{-4}$\\
\hline 
\end{tabular}
\end{table}
In this section, we evaluate the performance of our SRM-1D, Max-SLANR, and SCA schemes by Monte Carlo simulations. Furthermore, we develop a method based on ZF \cite{Xing2015To} to show the superiority of our schemes. Additionally, since in our proposed scheme, we take into account the estimation error of the direction angles of the eavesdroppers at the relay, we also consider the scenario relying on perfect estimation of the direction angle of eavesdroppers from the relay to arrive at a performance upper-bound of our schemes.

\begin{figure}[!ht]
  \centering
  \includegraphics[width=0.35\textwidth]{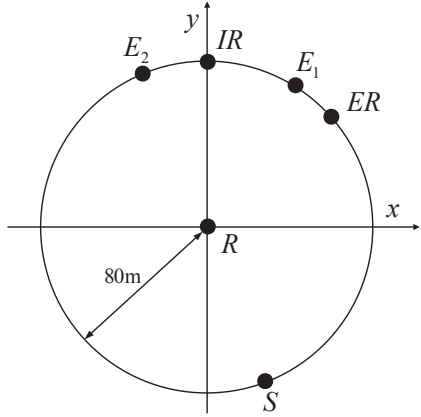}\\
  \caption{The location of source, relay, IR, ER and eavesdroppers.}\label{zuobiao}
\end{figure}

In the following, we denote by `SRM-1D-Perfect' and `SRM-1D-Robust' the SRM-1D method with perfect and imperfect knowledge of the                                   direction angles from the relay to the eavesdroppers, respectively, and represent by `SCA-Robust' and `Max-SLANR-Robust' our SCA and Max-SLANR schemes, respectively, when takeing into consideration the estimation error of direction angles at the relay. The simulation parameters are listed in Table II, unless otherwise stated.
The free-space path loss model used is defined as
\begin{align}\label{simu}
g_{mn}=\left(\frac{d_{mn}}{d_0}\right)^{-2},
\end{align}
where $g_{mn}$ and $d_{mn}$ denote the path loss and distance between node $m$ and node $n$, while $d_0$ is the reference distance, which is set to $10$m. The distances from relay to other nodes (source, IR, ER and eavesdroppers) are assumed to be the same, which are set to $80$ meters, i.e., $g_{mn}=\frac{1}{64}$. The direction angle of the source, IR, ER and eavesdroppers are $\theta_{sr}=-\frac{7\pi}{18}$, $\theta_{rd}=\frac{\pi}{2}$, $\theta_{rp}=\frac{\pi}{4}$ and $\{\theta_{re_1}, \theta_{re_2}\} =\{\frac{\pi}{3}, \frac{11\pi}{18}\}$, respectively. The location of the source, relay, IR, ER and eavesdroppers in the Cartesian coordinate system is shown in Fig.~\ref{zuobiao}.

\begin{figure}[!ht]
  \centering
  \includegraphics[width=0.48\textwidth]{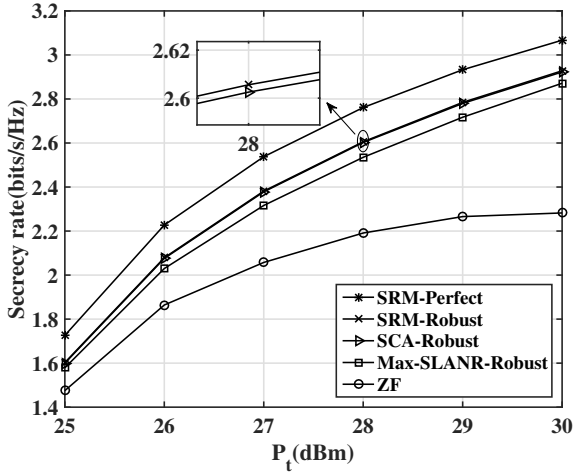}\\
  \caption{Secrecy rate versus the transmit power at relay for $N=6$, $M=2$, $P_s=30\mathrm{dBm}$, $P_{\mathrm{min}}=5\mathrm{dBm}$, $\Delta\theta_{\mathrm{max}}=6^\circ$.}\label{Sim_Pt_Rs}
\end{figure}

In Fig.~\ref{Sim_Pt_Rs}, we show the secrecy rate versus the total power $P_t$ at the relay. First, it can be observed that the secrecy rate grows upon increasing the transmit power $P_t$ at the relay for all cases. Second, compared to other schemes, SRM-1D-Perfect scheme has the best secrecy rate, because it has perfectly obtain the eavesdroppers' directional angle information. Third, the proposed SRM-1D-Robust slightly outperforms the proposed Max-SLANR-Robust arrangement. For example, when $P_t=30\mathrm{dBm}$, the secrecy rate of the Max-SLANR-Robust scheme is $0.05$bits/s/Hz lower than that of the SRM-1D-Robust scheme. This is because the SRM-1D-Robust scheme is the optimal solution, while the Max-SLANR-Robust scheme is a suboptimal solution. Fourth, the secrecy rate of SCA-Robust scheme is very close to that of the SRM-1D-Robust scheme. However, the complexity of the SCA-Robust scheme is much lower than that of the SRM-1D-Robust scheme. Finally, compared to the ZF scheme, our schemes provide significant performance improvement.
\begin{figure}[!ht]
  \centering
  \includegraphics[width=0.48\textwidth]{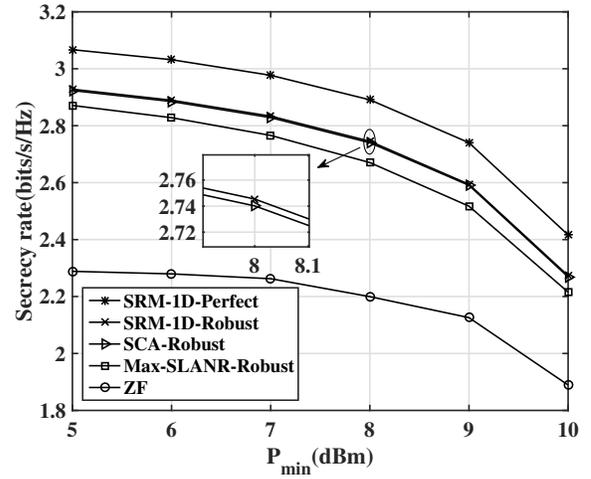}\\
  \caption{Secrecy rate versus the minimum energy required by the ER for $N=6$, $M=2$, $P_s=30\mathrm{dBm}$, $P_t=30\mathrm{dBm}$, $\Delta\theta_{\mathrm{max}}=6^\circ$.}\label{Sim_Pmin_Rs}
\end{figure}

Fig.~\ref{Sim_Pmin_Rs} shows the secrecy rate versus the minimum energy required by the ER for $N=6$, $P_s=30\mathrm{dBm}$, $P_t=30\mathrm{dBm}$, $\Delta\theta=6^\circ$. Naturally, the secrecy rate decreases with the increase of the minimum energy required by the ER for all cases. The reason behind this is that the more power is used for energy harvesting, the less power remains for secure communication, when the transmit power at the relay is fixed. Furthermore, when $P_{\mathrm{min}}<9\mathrm{dBm}$, it is observed that the secrecy rate decreases slowly. But when $P_{\mathrm{min}}>9\mathrm{dBm}$, the secrecy rate decreases rapidly. This is because the signal is transmitted over $80$m away from the relay, hence the power of the signal is only $12$dBm due to the path loss. Since the energy transfer efficiency is $\rho = 0.8$ and $P_{\mathrm{min}}>\mathrm{9dBm}$, the power of the transmit signal is mainly used for satisfying the energy harvesting constraint, which results in a rapid reduction of the secrecy rate for all the schemes.
\begin{figure}[!ht]
  \centering
  \includegraphics[width=0.48\textwidth]{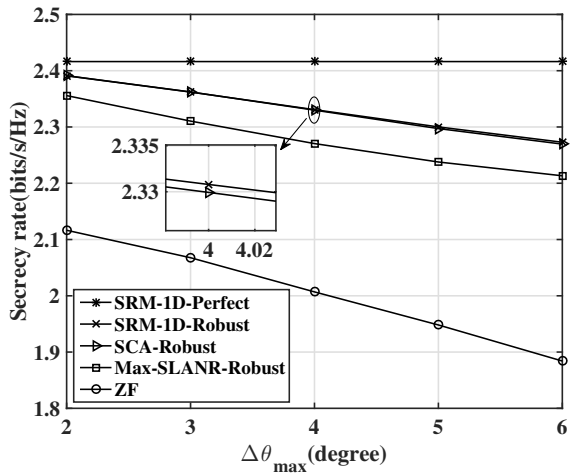}\\
  \caption{Secrecy rate versus the maximum estimate error angle for $N=6$, $M=2$, $P_s=30\mathrm{dBm}$, $P_t=30\mathrm{dBm}$, $P_{\mathrm{min}}=10\mathrm{dBm}$.}\label{Sim_Rs_theta}
\end{figure}

In Fig.~\ref{Sim_Rs_theta}, by fixing $P_t=30\mathrm{dBm}$ and $P_{\mathrm{min}}=10\mathrm{dBm}$, we investigate the effect of the maximum angular estimation error $\Delta\theta_{\mathrm{max}}$ of the eavesdroppers on the secrecy rate. The SRM-1D-Perfect curve remains constant for all $\Delta\theta_{\max}$ values and outperforms the other schemes due to the perfect knowledge of the directional angle. With the increase of $\Delta\theta_{\mathrm{max}}$, the secrecy rates achieved by robust schemes degrades slowly, because the proposed algorithms have considered the statistical property of the estimation error. As such, they are robust against the effects of estimation errors.

\begin{figure}[!ht]
  \centering
  \includegraphics[width=0.48\textwidth]{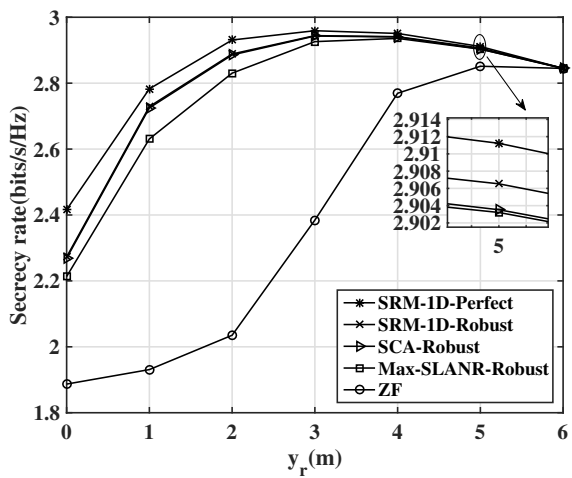}\\
  \caption{Secrecy rate versus the location of relay for $N=6$, $M=2$, $P_s=30\mathrm{dBm}$, $P_t=30\mathrm{dBm}$, $P_{\mathrm{min}}=10\mathrm{dBm}$, $\Delta\theta_{\mathrm{max}}=6^\circ$.}\label{location_relay_Rs}
\end{figure}

\begin{figure}[!ht]
  \centering
  \includegraphics[width=0.48\textwidth]{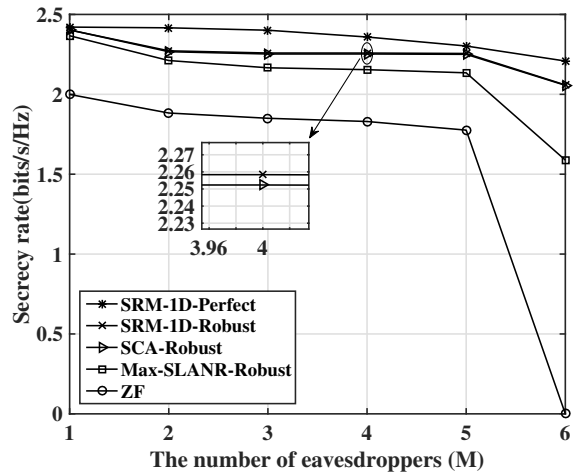}\\
 \caption{Secrecy rate versus the number of eavesdroppers for $N=6$, $\{\theta_{re_1}, \theta_{re_2},\theta_{re_3},\theta_{re_4},\theta_{re_5},\theta_{re_6}\} =\{\frac{\pi}{3}, \frac{11\pi}{18}, \frac{3\pi}{4}, \frac{\pi}{6}, \frac{2\pi}{3}, \frac{4\pi}{9}\}$, $P_s=30\mathrm{dBm}$, $P_t=30\mathrm{dBm}$, $P_{\mathrm{min}}=10\mathrm{dBm}$, $\Delta\theta_{\mathrm{max}}=6^\circ$.}\label{num_eav_Rs}
\end{figure}

Fig.~\ref{location_relay_Rs} shows the secrecy rate versus the location of the relay. We denote the coordinates of the relay by $(x_r, y_r)$. In this simulation, we fix $x_r$ to zero and assume that the relay moves vertically along the y-axis, starting from the origin towards the IR, while the locations of all the other nodes are fixed. As the relay moves, it becomes closer to the IR and farther from the source. We can see from Fig.~\ref{location_relay_Rs} that for all the schemes, the secrecy rate increases first and then decreases. The secrecy rate increases because the received signal-to-noise ratio (SNR) increases as the relay moves to the IR. The secrecy rate decreases because, as the relay continues approaching the IR, it is getting farther away from the source node, which decreases the SNR of the IR. We can observe that the optimal point is $y_r=30$m. Moreover, when $y_r=60$m, it is observed that the secrecy rates of all schemes converge to the same points, because when the relay has a low SNR, all these schemes have a similar secrecy rate performance.

\begin{figure}[!ht]
  \centering
  \includegraphics[width=0.48\textwidth]{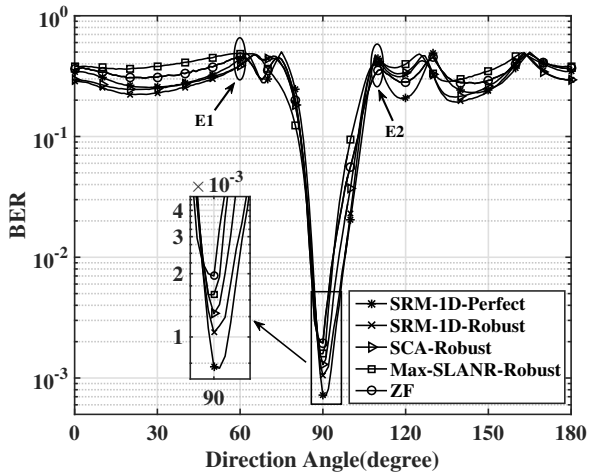}\\
  \caption{The performance of BER versus direction angle for $N=6$, $M=2$, $P_s=30\mathrm{dBm}$, $P_t=25\mathrm{dBm}$, $P_{\mathrm{min}}=5\mathrm{dBm}$, $\Delta\theta_{\mathrm{max}}=6^\circ$.}\label{Direction_BER}
\end{figure}

Fig.~\ref{num_eav_Rs} shows the secrecy rate versus the number of eavesdroppers. As can be seen from the Fig.~\ref{num_eav_Rs}, when $M = 2$ and $M = 6$, the secrecy rate of the proposed algorithm decreases rapidly. This is because the second and sixth eavesdroppers are located near the DR. Moreover, when $M=6$, the secrecy rate of the ZF scheme is $0$. This is because, the degrees of freedom of the relay is $6$ ($N = 6$), whereas the degrees of freedom at eavesdroppers are $6$ when $M=6$. Therefore, there is no degrees of freedom left for the DR.

Fig.~\ref{Direction_BER} studies the bit error rate (BER) versus the direction angle. We employ quadrature phase shift keying (QPSK) modulation. As seen from Fig.~\ref{Direction_BER}, the BER performance in the desired direction of $90^\circ$ is significantly better than in other directions for all cases. Observe from Fig.~\ref{Direction_BER}, that in the vicinity of the two eavesdroppers' directions, the BER performance is poor, since the signals in these two directions are contaminated by the AN. Thus the eavesdroppers cannot successfully receive the information destined to the IR.

\section{Conclusion}\label{conclusion}
In this paper, we investigated secure wireless information and power transfer based on DM in AF relay networks. Specifically, the robust information beamforming matrix and the AN covariance matrix were designed based on the SRM-1D scheme and on the Max-SLANR scheme. To solve the optimization problem of SRM, we proposed a twin-level optimization method that includes a 1D search and the SDR technique. Furthermore, we proposed a suboptimal solution for the SRM problem, which was based on the Max-SLANR criterion. Finally, we formulated a SRM problem, which was transformed into a SOCP problem, and solved by a low-complexity SCA method. Simulation results show that the performance of the SCA scheme is very close to that of the SRM-1D scheme in terms of its secrecy rate and bit error rate (BER), and compared to the ZF scheme, the SCA scheme and Max-SLANR schemes provide a significant performance improvement.

\appendices

\section{Derivation of $H_{re_m}(p,q)$}
$H_{re_m}(p,q)$ can be expressed as
\begin{align}\label{APP1_1}
&H_{re_m}(p,q)=\mathbb{E}\left[\mathbf{h}_p(\hat{\theta}_{re_m}+\Delta\theta_{re_m})\mathbf{h}_q^H(\hat{\theta}_{re_m}+\Delta\theta_{re_m})\right]\nonumber\\
&=\frac{g_{re_m}}{N}\int_{-\Delta\theta_{max}}^{\Delta\theta_{max}}e^{j\alpha_{pq}\cos(\hat{\theta}_{re_m}+\Delta\theta_{re_m})}f\left(\Delta\theta_{re_m}\right)d(\Delta\theta_{re_m})\nonumber\\
&=\frac{g_{re_m}}{N}\int_{-\Delta\theta_{max}}^{\Delta\theta_{max}}\text{exp}\Big\{j\alpha_{pq}\big(\cos(\hat{\theta}_{re_m})\cos(\Delta\theta_{re_m})\nonumber\\
&-\sin(\hat{\theta}_{re_m})\sin(\Delta\theta_{re_m})\big)\Big\}f\left(\Delta\theta_{re_m}\right)d(\Delta\theta_{re_m}),
\end{align}
where $\alpha_{pq}=\frac{2\pi(q-p)l}{\lambda}$. Assuming $\Delta\theta_{max}$ is a small value near zero, we get the following approximate expression using second-order Taylor expansion
\begin{align}\label{APP1_2}
\sin(\Delta\theta_{re_m})\approx&\Delta\theta_{re_m}\nonumber\\
\cos(\Delta\theta_{re_m})\approx& 1-\frac{\Delta\theta_{re_m}^2}{2}.
\end{align}
Substituting $(\ref{APP1_2})$ into $(\ref{APP1_1})$ yields
\begin{align}\label{APP1_3}
&H_{re_m}(p,q)=\hat{H}_{re_m}(p,q)\int_{-\Delta\theta_{max}}^{\Delta\theta_{max}}\nonumber\\
&e^{-j\alpha_{pq}\left(\cos(\hat{\theta}_{re_m})\frac{\Delta\theta_{re_m}^2}{2}+\sin(\hat{\theta}_{re_m})\Delta\theta_{re_m}\right)}f\left(\Delta\theta_{re_m}\right)d(\Delta\theta_{re_m})\nonumber\\
&=\hat{H}_{re_m}(p,q)\int_{-\Delta\theta_{max}}^{\Delta\theta_{max}}\left[\cos\left(\xi_{p,q}^m\right)-j\sin\left(\xi_{p,q}^m\right)\right]\nonumber\\
&~~~~f\left(\Delta\theta_{re_m}\right)d(\Delta\theta_{re_m})\nonumber\\
&=\Gamma_{1_m}(p,q)-j\Gamma_{2_m}(p,q),
\end{align}
where
\begin{align}\label{APP1_4}
\hat{H}_{re_m}(p,q)=\frac{g_{re_m}}{N}e^{j\alpha_{pq}\cos(\hat{\theta}_{re_m})}
\end{align}
represents the $p$-th row and $q$-th column entry of $\mathbf{h}(\hat{\theta}_{re_m})\mathbf{h}^H(\hat{\theta}_{re_m})$, and
\begin{subequations}\label{APP1_5:1}
\begin{align}
&\xi_{p,q}^m=\alpha_{pq}\left(\cos(\hat{\theta}_{re_m})\frac{\Delta\theta_{re_m}^2}{2}+\sin(\hat{\theta}_{re_m})\Delta\theta_{re_m}\right),\\
&\Gamma_{1_m}(p,q)=\nonumber\\
&\hat{H}_{re_m}(p,q)\int_{-\Delta\theta_{max}}^{\Delta\theta_{max}}\cos\left(\xi_{p,q}^m\right)f\left(\Delta\theta_{re_m}\right)d(\Delta\theta_{re_m}),\\
&\Gamma_{2_m}(p,q)=\nonumber\\
&\hat{H}_{re_m}(p,q)\int_{-\Delta\theta_{max}}^{\Delta\theta_{max}}\sin\left(\xi_{p,q}^m\right)f\left(\Delta\theta_{re_m}\right)d(\Delta\theta_{re_m}).
\end{align}
\end{subequations}
Now the task is to derive the analytic expression of $\Gamma_{1_m}(p,q)$ and $\Gamma_{2_m}(p,q)$. To this end, we first expand the trigonometric function into the following form
\begin{align}\label{APP1_6}
&\cos\left(\xi_{p,q}^m\right)=\\
&\cos\left(\alpha_{pq}\cos(\hat{\theta}_{re_m})\frac{\Delta\theta_{re_m}^2}{2}+\alpha_{pq}\sin(\hat{\theta}_{re_m})\Delta\theta_{re_m}\right)\nonumber\\
&=\cos\left(\alpha_{pq}\cos(\hat{\theta}_{re_m})\frac{\Delta\theta_{re_m}^2}{2}\right)\cos\left(\alpha_{pq}\sin(\hat{\theta}_{re_m})\Delta\theta_{re_m}\right)\nonumber\\
&-\sin\left(\alpha_{pq}\cos(\hat{\theta}_{re_m})\frac{\Delta\theta_{re_m}^2}{2}\right)\sin\left(\alpha_{pq}\sin(\hat{\theta}_{re_m})\Delta\theta_{re_m}\right).\nonumber
\end{align}
Then using the second-order Taylor series to approximate each term, we can rewrite $(\ref{APP1_6})$ as
\begin{align}\label{APP1_7}
&\cos\left(\xi_{p,q}^m\right)=\left(1-\frac{\alpha_{pq}^2\cos^2(\hat{\theta}_{re_m})\Delta\theta_{re_m}^4}{8}\right)\times\nonumber\\
&\left(1-\frac{\alpha_{pq}^2\sin^2(\hat{\theta}_{re_m})\Delta\theta_{re_m}^2}{2}\right)\nonumber\\
&-\left(\alpha_{pq}\cos(\hat{\theta}_{re_m})\frac{\Delta\theta_{re_m}^2}{2}\right)\left(\alpha_{pq}\sin(\hat{\theta}_{re_m})\Delta\theta_{re_m}\right)\nonumber\\
&\approx\left(1-\frac{1}{2}\alpha_{pq}^2\sin^2(\hat{\theta}_{re_m})\Delta\theta_{re_m}^2\right)-\nonumber\\
&\left(\alpha_{pq}\cos(\hat{\theta}_{re_m})\frac{\Delta\theta_{re_m}^2}{2}\right)\left(\alpha_{pq}\sin(\hat{\theta}_{re_m})\Delta\theta_{re_m}\right).
\end{align}

\newcounter{mytempeqncnt}
\begin{figure*}[!t]
\normalsize
\setcounter{mytempeqncnt}{\value{equation}}
\begin{align}\label{APP1_8}
\Gamma_{1_m}(p,q)&\approx\hat{H}_{re_m}(p,q)\int_{-\Delta\theta_{max}}^{\Delta\theta_{max}}\left(1-\frac{1}{2}\alpha_{pq}^2\sin^2(\hat{\theta}_{re_m})\Delta\theta_{re_m}^2\right)
f\left(\Delta\theta_{re_m}\right)d(\Delta\theta_{re_m})\nonumber\\
&\overset{(a)}{=}\frac{\hat{H}_{re_m}(p,q)}{K_{e_m}}\left(\mathrm{erf}(\frac{\Delta\theta_{max}}{\sqrt{2}\sigma_{\theta}})-\alpha_{pq}^2\sin^2(\hat{\theta}_{re_m})\sigma_{\theta}^2\left(\frac{1}{2}\mathrm{erf}(\frac{\Delta\theta_{max}}{\sqrt{2}\sigma_{\theta}})-\frac{\Delta\theta_{max}}{\sqrt{2\pi}\sigma_{\theta}}e^{-\frac{1}{2\sigma_{\theta}^2}\Delta\theta_{max}^2}\right)\right),
\tag{64}
\end{align}
\begin{align}\label{APP1_11}
\Gamma_{2_m}(p,q)&\approx\hat{H}_{re_m}(p,q)\int_{-\Delta\theta_{max}}^{\Delta\theta_{max}}\left(\frac{1}{2}\alpha_{pq}\cos(\hat{\theta}_{re_m})\Delta\theta_{re_m}^2\right)
f\left(\Delta\theta_{re_m}\right)d(\Delta\theta_{re_m})\nonumber\\
&=\frac{\hat{H}_{re_m}(p,q)\alpha_{pq}\cos(\hat{\theta}_{re_m})\sigma_{\theta}^2}{K_{e_m}}\left(\frac{1}{2}\mathrm{erf}(\frac{\Delta\theta_{max}}{\sqrt{2}\sigma_{\theta}})-\frac{\Delta\theta_{max}}{\sqrt{2\pi}\sigma_{\theta}}e^{-\frac{1}{2\sigma_{\theta}^2}\Delta\theta_{max}^2}\right).
\tag{67}
\end{align}
\begin{align}\label{APP2_2}
&L(\tilde{\mathbf{W}},\mathbf{\Omega},\mu,\nu_m,\eta,\zeta,\mathbf{S}_1,\mathbf{S}_2)=\nonumber\\
&P_s\mathrm{Tr}(\mathbf{C}_1\tilde{\mathbf{W}})+\sigma^2\mathrm{Tr}(\tilde{\mathbf{W}})+\mu\left(-P_s\mathrm{Tr}(\mathbf{A}_1\tilde{\mathbf{W}})+\sigma^2\phi(\beta)\mathrm{Tr}(\mathbf{A}_2\tilde{\mathbf{W}})+\phi(\beta)\mathbf{h}^H(\theta_{rd})\mathbf{\Omega}\mathbf{h}(\theta_{rd})+\phi(\beta)\sigma^2\right)\nonumber\\
&+\sum_{m=1}^M\nu_m\left(P_s\mathrm{Tr}(\mathbf{B}_{1_m}\tilde{\mathbf{W}})-\beta\sigma^2\mathrm{Tr}(\mathbf{B}_{2_m}\tilde{\mathbf{W}})-\beta\mathrm{Tr}(\bar{\mathbf{H}}_{re_m}\mathbf{\Omega})-\beta\sigma^2\right)+\eta\left(P_s\mathrm{Tr}(\mathbf{C}_1\tilde{\mathbf{W}})+\sigma^2\mathrm{Tr}(\tilde{\mathbf{W}})+\mathrm{Tr}(\mathbf{\Omega})-P_t\right)\nonumber\\
&-\zeta\left(P_s\mathrm{Tr}(\mathbf{D}_1\tilde{\mathbf{W}})+\sigma^2\mathrm{Tr}(\mathbf{D}_2\tilde{\mathbf{W}})+\mathbf{h}^H(\theta_{rp})\mathbf{\Omega}\mathbf{h}(\theta_{rp})-\frac{P_{min}}{\rho}\right)-\mathrm{Tr}(\mathbf{S}_1\tilde{\mathbf{W}})-\mathrm{Tr}(\mathbf{S}_2\mathbf{\Omega}).
\tag{69}
\end{align}
\setcounter{equation}{\value{mytempeqncnt}}
\hrulefill
\vspace*{4pt}
\end{figure*}

\setcounter{equation}{64}
Since the last term in  $(\ref{APP1_7})$ is an odd function with respect to $\Delta\theta_{re_m}$, hence we have (\ref{APP1_8}) at the top of the next page. Note that step (a) in $(\ref{APP1_8})$ results from the following equation \cite{Gradshteyn2007}
\begin{align}\label{APP1_9}
\int_0^xt^2e^{-q^2t^2}dt=\frac{1}{2q^3}\left(\frac{\sqrt{\pi}}{2}\mathrm{erf}(qx)-qxe^{-q^2x^2}\right),
\end{align}
where $\mathrm{erf}(x)$ represents the error function defined as
\begin{align}\label{APP1_10}
\mathrm{erf}(x)=\frac{2}{\sqrt{\pi}}\int_0^xe^{-t^2}dt.
\end{align}
Using a similar method as above, $\Gamma_{2_m}(p,q)$ can be approximated as (\ref{APP1_11}) at the top of the page. Combining $(\ref{APP1_3})$, $(\ref{APP1_8})$ and $(\ref{APP1_11})$, we obtain the analytic expression of $H_{re_m}(p,q)$ and the proof is completed.\hfill$\blacksquare$
\setcounter{equation}{67}

\section{Proof of Lemma 1}
The optimization problem $(\ref{per10})$ can be rewritten as
\begin{align}\label{APP2_1}
&\min_{{\tilde{\mathbf{W}},\mathbf{\Omega}}}~ P_s\mathrm{Tr}(\mathbf{C}_1\tilde{\mathbf{W}})+\sigma^2\mathrm{Tr}(\tilde{\mathbf{W}})\nonumber\\
&\mathrm{s.t.}~-P_s\mathrm{Tr}(\mathbf{A}_1\tilde{\mathbf{W}})+\sigma^2\phi(\beta)\mathrm{Tr}(\mathbf{A}_2\tilde{\mathbf{W}})+\nonumber\\
&~~~~~\phi(\beta)\mathbf{h}^H(\theta_{rd})\mathbf{\Omega}\mathbf{h}(\theta_{rd})+\phi(\beta)\sigma^2\leq 0,\nonumber\\
&~~~~~(\ref{per7:1b}),(\ref{per7:1c}),(\ref{per7:1d}),\tilde{\mathbf{W}}\succeq\mathbf{0},\mathbf{\Omega}\succeq\mathbf{0}.
\end{align}
Since the problem  $(\ref{APP2_1})$ satisfies Slater's constraint qualification \cite{Boyd}, its objective function and constraints are convex, and the optimal solution must satisfy KKT conditions. The Lagrangian of $(\ref{APP2_1})$ is given in (\ref{APP2_2}) at the top of the page, where $\mu,\nu_m,\eta,\zeta,\mathbf{S}_1$ and $\mathbf{S}_2$ denote the dual variables associated with the constraint in  $(\ref{APP2_1})$. Let $\tilde{\mathbf{W}}^\star,\mathbf{\Omega}^\star$ be the optimal primal variables and $\mu^\star,\nu_m^\star,\eta^\star,\zeta^\star,\mathbf{S}_1^\star,\mathbf{S}_2^\star$ be the dual variables. The Karush-Kuhn-Tucker (KKT) conditions related to the proof are given as follows
\setcounter{equation}{69}
\begin{subequations}\label{APP2_3:1}
\begin{align}
&\mathbf{S}_1^\star=P_s\mathbf{C}_1+\sigma^2\mathbf{I}_{N^2}-\mu^\star P_s\mathbf{A}_1+\mu^\star\sigma^2\phi(\beta)\mathbf{A}_2+\nonumber\\
&~~~~P_s\sum_{m=1}^M\nu_m^\star\mathbf{B}_{1_m}-\beta\sigma^2\sum_{m=1}^M\nu_m^\star\mathbf{B}_{2_m}+\eta^\star P_s\mathbf{C}_1+\nonumber\\
&~~~~\eta^\star\sigma^2\mathbf{I}_{N^2}-\zeta^\star P_s\mathbf{D}_1-\zeta^\star\sigma^2\mathbf{D}_2,\label{APP2_3:1a}\\
&\mathbf{S}_2^\star=\mu^\star\phi(\beta)\mathbf{h}(\theta_{rd})\mathbf{h}^H(\theta_{rd})-\beta\sum_{m=1}^M\nu_m^\star\mathbf{H}_{re_m}+\nonumber\\
&~~~~\eta^\star\mathbf{I}_N-\zeta^\star\mathbf{h}(\theta_{rp})\mathbf{h}^H(\theta_{rp}),\label{APP2_3:1b}\\
&\mathbf{S}_1^\star\tilde{\mathbf{W}}^\star=\mathbf{0},\mathbf{S}_2^\star\mathbf{\Omega}^\star=\mathbf{0},\mathbf{S}_1^\star\succeq \mathbf{0},\mathbf{S}_2^\star\succeq \mathbf{0},\tilde{\mathbf{W}}\succeq \mathbf{0}.\label{APP2_3:1c}
\end{align}
\end{subequations}
From $(\ref{APP2_3:1b})$, we arrive at:
\begin{align}\label{APP2_4}
&\left(P_s\mathbf{h}^*(\theta_{sr})\mathbf{h}^T(\theta_{sr})+\sigma^2\mathbf{I}_N\right)\otimes\mathbf{S}_2^\star=\mu^\star P_s\phi(\beta)\mathbf{A}_1+\nonumber\\
&\mu^\star\sigma^2\phi(\beta)\mathbf{A}_2-\beta P_s\sum_{m=1}^M\nu_m^\star\mathbf{B}_{1_m}-\beta\sigma^2\sum_{m=1}^M\nu_m^\star\mathbf{B}_{2_m}+\nonumber\\
&\eta^\star P_s\mathbf{C}_1+\eta^\star\sigma^2\mathbf{I}_{N^2}-\zeta^\star P_s\mathbf{D}_1-\zeta^\star\sigma^2\mathbf{D}_2.
\end{align}
Substituting $(\ref{APP2_4})$ into $(\ref{APP2_3:1a})$, we have
\begin{align}\label{APP2_5}
\mathbf{S}_1^\star+\mu^\star P_s\left[\phi(\beta)+1\right]\mathbf{A}_1=\mathbf{\Xi},
\end{align}
where
\begin{align}\label{APP2_6}
\mathbf{\Xi}=&\sigma^2\mathbf{I}_{N^2}+P_s\mathbf{C}_1+P_s(\beta+1)\sum_{m=1}^M\nu_m^\star\mathbf{B}_{1_m}+\nonumber\\
&\left[P_s\mathbf{h}^*(\theta_{sr})\mathbf{h}^T(\theta_{sr})+\sigma^2\mathbf{I}_N\right]\otimes\mathbf{S}_2^\star.
\end{align}
Let us multiply both sides of $(\ref{APP2_5})$ by $\tilde{\mathbf{W}}^\star$ and substitute the first term of $(\ref{APP2_3:1c})$ into the resultant equation, yielding:
\begin{align}\label{APP2_7}
 \mu^\star P_s\left(\phi(\beta)+1\right)\mathbf{A}_1\tilde{\mathbf{W}}^\star=\mathbf{\Xi}\tilde{\mathbf{W}}^\star.
\end{align}
Since $\left(P_s\mathbf{h}^*(\theta_{sr})\mathbf{h}^T(\theta_{sr})+\sigma^2\mathbf{I}_N\right)$ is a Hermitian positive definite matrix, and $\mathbf{S}_2^\star\succeq\mathbf{0}$, according to the fact that \cite{Golub1996Matrix}
\begin{align}\label{APP2_8}
 \lambda(\mathbf{X}\otimes\mathbf{Y})=\{\mu_i\gamma_j,\mu_i\in\lambda(\mathbf{X}),\gamma_j\in\lambda(\mathbf{Y})\},
\end{align}
we have
\begin{align}\label{APP2_8_1}
\left[P_s\mathbf{h}^*(\theta_{sr})\mathbf{h}^T(\theta_{sr})+\sigma^2\mathbf{I}_N\right]\otimes\mathbf{S}_2^\star\succeq\mathbf{0},
\end{align}
where $\lambda(\bullet)$ denotes the set of  matrix eigenvalues.
Upon observing that the first term in  $(\ref{APP2_6})$ is the identity matrix and the other three terms are Hermitian semidefinite matrices,
we have $\mathbf{\Xi}\succ\mathbf{0}$. By exploiting the fact that $\mathrm{rank}(\mathbf{AB})\leq \mathrm{min}\{\mathrm{rank}(\mathbf{A}),\mathrm{rank}(\mathbf{B})\}$, we obtain
\begin{align}\label{APP2_9}
\mathrm{rank}(\mathbf{\Xi}\tilde{\mathbf{W}}^\star)=\mathrm{rank}(\tilde{\mathbf{W}}^\star)\leq\mathrm{rank}( \mu^\star P_s\left(\phi(\beta)+1\right)\mathbf{A}_1).
\end{align}
Combining $(\ref{APP2_7})$ with $(\ref{APP2_9})$ and considering that $\mathbf{A}_1$ is a rank-one matrix, we conclude that $\mathrm{rank}(\tilde{\mathbf{W}}^\star)=1$ for $\mu^\star> 0$, $\mathrm{rank}(\tilde{\mathbf{W}}^\star)=0$ for $\mu^\star= 0$. However, $\tilde{\mathbf{W}}^\star=\mathbf{0}$ corresponds to no signal transmission. Hence we can conclude that $\mathrm{rank}(\tilde{\mathbf{W}}^\star)=1$, and the proof is completed.\hfill$\blacksquare$

\section{Proof of Lemma 2}
Since the problem  $(\ref{SL5})$  satisfies Slater's constraint qualification \cite{Boyd}, its objective function and constraints are convex, and the optimal solution must satisfy KKT conditions. The Lagrangian associated with the problem $(\ref{SL5})$ is given by
\begin{align}\label{APP4_1}
&L(\tilde{\mathbf{W}},\mathbf{\Omega},\mu,\eta,\zeta,\mathbf{S}_1,\mathbf{S}_2)=P_s\mathrm{Tr}(\mathbf{C}_1\tilde{\mathbf{W}})+\sigma^2\mathrm{Tr}(\tilde{\mathbf{W}})\nonumber\\
&+\mu\Big(-P_s\mathrm{Tr}(\mathbf{A}_1\tilde{\mathbf{W}})+\phi\sum_{m=1}^MP_s\mathrm{Tr}(\mathbf{B}_{1_m}\tilde{\mathbf{W}})\nonumber\\
&+\phi\mathbf{h}^H(\theta_{rd})\mathbf{\Omega}\mathbf{h}(\theta_{rd})+\phi\sigma^2\mathrm{Tr}(\mathbf{A}_2\tilde{\mathbf{W}})+\phi\sigma^2\Big)\nonumber\\
&+\eta\left(P_s\mathrm{Tr}(\mathbf{C}_1\tilde{\mathbf{W}})+\sigma^2\mathrm{Tr}(\tilde{\mathbf{W}})+\mathrm{Tr}(\mathbf{\Omega})-P_t\right)\nonumber\\
&-\zeta\Big(P_s\mathrm{Tr}(\mathbf{D}_1\tilde{\mathbf{W}})+\sigma^2\mathrm{Tr}(\mathbf{D}_2\tilde{\mathbf{W}})+\mathbf{h}^H(\theta_{rp})\mathbf{\Omega}\mathbf{h}(\theta_{rp})\nonumber\\
&-\frac{P_{min}}{\rho}\Big)-\mathrm{Tr}(\mathbf{S}_1\tilde{\mathbf{W}})-\mathrm{Tr}(\mathbf{S}_1\tilde{\mathbf{W}})-\mathrm{Tr}(\mathbf{S}_2\mathbf{\Omega}),
\end{align}
where $\mu,\eta,\zeta,\mathbf{S}_1$ and $\mathbf{S}_2$ are dual variables associated with the constraint in $(\ref{SL5})$, while $\tilde{\mathbf{W}}^\star,\mathbf{\Omega}^\star$ are the optimal primal variables and $\mu^\star,\eta^\star,\zeta^\star,\mathbf{S}_1^\star,\mathbf{S}_2^\star$ are dual variables. The KKT conditions that are relevant to the proof are given by
\begin{subequations}\label{APP4_2:1}
\begin{align}
&\mathbf{S}_1^\star=P_s\mathbf{C}_1-\mu^\star P_s\mathbf{A}_1+\mu^\star\phi\sigma^2\mathbf{A}_2+\mu^\star\phi P_s\sum_{m=1}^M\mathbf{B}_{1_m}+\nonumber\\
&\sigma^2\mathbf{I}_{N^2}+\eta^\star P_s\mathbf{C}_1+\eta^\star\sigma^2\mathbf{I}_{N^2}-\zeta^\star P_s\mathbf{D}_1-\zeta^\star\sigma^2\mathbf{D}_2,\label{APP4_2:1a}\\
&\mathbf{S}_2^\star=\mu^\star\phi\mathbf{h}(\theta_{rd})\mathbf{h}^H(\theta_{rd})+\eta^\star\mathbf{I}_N-\zeta^\star\mathbf{h}(\theta_{rp})\mathbf{h}^H(\theta_{rp}),\label{APP4_2:1b}\\
&\mathbf{S}_1^\star\tilde{\mathbf{W}}^\star=\mathbf{0},\mathbf{S}_2^\star\mathbf{\Omega}^\star=\mathbf{0},\mathbf{S}_1^\star\succeq \mathbf{0},\mathbf{S}_2^\star\succeq \mathbf{0},\tilde{\mathbf{W}}\succeq \mathbf{0}.\label{APP4_2:1c}
\end{align}
\end{subequations}
From $(\ref{APP4_2:1b})$, we get
\begin{align}\label{APP4_3}
&\left[P_s\mathbf{h}^*(\theta_{sr})\mathbf{h}^T(\theta_{sr})+\sigma^2\mathbf{I}_N\right]\otimes\mathbf{S}_2^\star=\mu^\star P_s\phi\mathbf{A}_1+\\
&\mu^\star\sigma^2\phi\mathbf{A}_2
+\eta^\star P_s\mathbf{C}_1+\eta^\star\sigma^2\mathbf{I}_{N^2}-\zeta^\star P_s\mathbf{D}_1-\zeta^\star\sigma^2\mathbf{D}_2.\nonumber
\end{align}
Substituting $(\ref{APP4_3})$ into $(\ref{APP4_2:1a})$, we have
\begin{align}\label{APP4_4}
\mathbf{S}_1^\star+\mu^\star P_s\left(1+\phi\right)\mathbf{A}_1=\mathbf{\Xi},
\end{align}
where
\begin{align}\label{APP4_5}
\mathbf{\Xi}=&\sigma^2\mathbf{I}_{N^2}+P_s\mathbf{C}_1+\mu^\star\phi P_s\sum_{m=1}^M\mathbf{B}_{1_m}+\nonumber\\
&\left[P_s\mathbf{h}^*(\theta_{sr})\mathbf{h}^T(\theta_{sr})+\sigma^2\mathbf{I}_N\right]\otimes\mathbf{S}_2^\star.
\end{align}
It is clear from $(\ref{APP4_5})$ that $\mathbf{\Xi}$ is a Hermitian positive definite matrix. The remaining steps of the proof are similar to the \emph{Lemma 1} and is omitted here. The proof is completed.\hfill$\blacksquare$
\bibliographystyle{IEEEtran}
\bibliography{IEEEfull,cooperative_bib}

\begin{thebibliography}{10}
\providecommand{\url}[1]{#1}
\csname url@samestyle\endcsname
\providecommand{\newblock}{\relax}
\providecommand{\bibinfo}[2]{#2}
\providecommand{\BIBentrySTDinterwordspacing}{\spaceskip=0pt\relax}
\providecommand{\BIBentryALTinterwordstretchfactor}{4}
\providecommand{\BIBentryALTinterwordspacing}{\spaceskip=\fontdimen2\font plus
\BIBentryALTinterwordstretchfactor\fontdimen3\font minus
  \fontdimen4\font\relax}
\providecommand{\BIBforeignlanguage}[2]{{%
\expandafter\ifx\csname l@#1\endcsname\relax
\typeout{** WARNING: IEEEtran.bst: No hyphenation pattern has been}%
\typeout{** loaded for the language `#1'. Using the pattern for}%
\typeout{** the default language instead.}%
\else
\language=\csname l@#1\endcsname
\fi
#2}}
\providecommand{\BIBdecl}{\relax}
\BIBdecl

\bibitem{Wu2018Spectral}
Q.~Wu, W.~Chen, D.~W.~K. Ng, and R.~Schober, ``Spectral and energy efficient
  wireless powered {IoT} networks: Noma or tdma?'' \emph{IEEE Trans. Veh.
  Technol.}, vol.~PP, no.~99, pp. 1--1, 2018.

\bibitem{Wu2016Joint}
Q.~Wu, M.~Tao, and W.~Chen, ``Joint {T}x/{R}x energy-efficient scheduling in
  multi-radio wireless networks: A divide-and-conquer approach,'' \emph{IEEE
  Trans. Wireless Commun.}, vol.~15, no.~4, pp. 2727--2740, Apr. 2016.

\bibitem{HSonjoint2014}
H.~Son and B.~Clerckx, ``Joint beamforming design for multi-user wireless
  information and power transfer,'' \emph{IEEE Trans. Wireless Commun.},
  vol.~13, no.~11, pp. 6397--6409, Nov 2014.

\bibitem{Chu2016}
Z.~Chu, Z.~Zhu, M.~Johnston, and S.~Y.~L. Goff, ``Simultaneous wireless
  information power transfer for {MISO} secrecy channel,'' \emph{IEEE Trans.
  Veh. Technol.}, vol.~65, no.~9, pp. 6913--6925, Sep. 2016.

\bibitem{Zhang2016}
H.~Zhang, Y.~Huang, C.~Li, and L.~Yang, ``Secure beamforming design for {SWIPT}
  in {MISO} broadcast channel with confidential messages and external
  eavesdroppers,'' \emph{IEEE Trans. Wireless Commun.}, vol.~15, no.~11, pp.
  7807--7819, Nov. 2016.

\bibitem{Ng2013}
D.~W.~K. Ng, E.~S. Lo, and R.~Schober, ``Robust beamforming for secure
  communication in systems with wireless information and power transfer,''
  \emph{IEEE Trans. Wireless Commun.}, vol.~13, no.~8, pp. 4599--4615, Aug.
  2013.

\bibitem{Wu2015}
W.~Wu and B.~Wang, ``Robust secrecy beamforming for wireless information and
  power transfer in multiuser {MISO} communication system,'' \emph{Eurasip
  Journal on Wireless Communications Networking}, vol. 2015, no.~1, pp.
  161--171, Jun. 2015.

\bibitem{Wu2016En}
Q.~Wu, G.~Li, W.~Chen, and D.~W.~K. Ng, ``Energy-efficient small cell with
  spectrum-power trading,'' \emph{IEEE J. Sel. Areas Commun.}, vol.~34, no.~12,
  pp. 3394--3408, Dec. 2016.

\bibitem{Khandaker2015}
M.~R.~A. Khandaker and K.~K. Wong, ``Robust secrecy beamforming with
  energy-harvesting eavesdroppers,'' \emph{IEEE Wireless Commun. Lett.},
  vol.~4, no.~1, pp. 10--13, Feb. 2015.

\bibitem{ArtificialMEI2017}
W.~Mei, Z.~Chen, and J.~Fang, ``Artificial noise aided energy efficiency
  optimization in {MIMOME} system with {SWIPT},'' \emph{IEEE Commun. Lett.},
  vol.~21, no.~8, pp. 1795--1798, Aug 2017.

\bibitem{Xun2013}
Z.~Xun, Z.~Rui, and C.~K. Ho, ``Wireless information and power transfer:
  Architecture design and rate-energy tradeoff,'' \emph{IEEE Trans. Commun.},
  vol.~61, no.~11, pp. 4754--4767, Nov. 2013.

\bibitem{Zhang2013MIMO}
R.~Zhang and C.~K. Ho, ``{MIMO} broadcasting for simultaneous wireless
  information and power transfer,'' \emph{IEEE Trans. Wireless Commun.},
  vol.~12, no.~5, pp. 1989--2001, May. 2013.

\bibitem{WU2017}
Q.~Wu, G.~Y. Li, W.~Chen, D.~W.~K. Ng, and R.~Schober, ``An overview of
  sustainable green {5G} networks,'' \emph{IEEE Wireless Commun.}, vol.~24,
  no.~4, pp. 72--80, Aug 2017.

\bibitem{Wu2016Energy}
Q.~Wu, M.~Tao, D.~W.~K. Ng, W.~Chen, and R.~Schober, ``Energy-efficient
  resource allocation for wireless powered communication networks,'' \emph{IEEE
  Trans. Wireless Commun.}, vol.~15, no.~3, pp. 2312--2327, March 2016.

\bibitem{Wu2016User}
Q.~Wu, W.~Chen, D.~W.~K. Ng, J.~Li, and R.~Schober, ``User-centric energy
  efficiency maximization for wireless powered communications,'' \emph{IEEE
  Trans. Wireless Commun.}, vol.~15, no.~10, pp. 6898--6912, Oct. 2016.

\bibitem{Wu2017Energy}
Q.~Wu, G.~Li, W.~Chen, and D.~W.~K. Ng, ``Energy-efficient d2d overlaying
  communications with spectrum-power trading,'' \emph{IEEE Trans. Wireless
  Commun.}, vol.~16, no.~7, pp. 4404--4419, Jul 2017.

\bibitem{Chen2015Secrecy}
X.~Chen, D.~W.~K. Ng, and H.~H. Chen, ``Secrecy wireless information and power
  transfer: challenges and opportunities,'' \emph{IEEE Wireless Commun.},
  vol.~23, no.~2, pp. 54--61, Apr. 2015.

\bibitem{Zou2015A}
Y.~Zou, J.~Zhu, X.~Wang, and L.~Hanzo, ``A survey on wireless security:
  Technical challenges, recent advances, and future trends,'' \emph{Proc.
  IEEE}, vol. 104, no.~9, pp. 1727--1765, Sep. 2015.

\bibitem{Zou2014Relay}
Y.~Zou, B.~Champagne, W.~P. Zhu, and L.~Hanzo, ``Relay-selection improves the
  security-reliability trade-off in cognitive radio systems,'' \emph{IEEE
  Trans. Commu.}, vol.~63, no.~1, pp. 215--228, Jan. 2014.

\bibitem{RelayBC}
Y.~Huang and B.~Clerckx, ``Relaying strategies for wireless-powered mimo relay
  networks,'' \emph{IEEE Trans. Wireless Commun.}, vol.~15, no.~9, pp.
  6033--6047, Sep 2016.

\bibitem{KimNon}
P.~Liu, S.~Gazor, I.~M. Kim, and D.~I. Kim, ``Noncoherent relaying in energy
  harvesting communication systems,'' \emph{IEEE Trans. Wireless Commun.},
  vol.~14, no.~12, pp. 6940--6954, Dec 2015.

\bibitem{Li2014Secure}
Q.~Li, Q.~Zhang, and J.~Qin, ``Secure relay beamforming for simultaneous
  wireless information and power transfer in nonregenerative relay networks,''
  \emph{IEEE Trans. Veh. Technol.}, vol.~63, no.~5, pp. 2462--2467, Jun. 2014.

\bibitem{Salem2016Physical}
A.~Salem, K.~A. Hamdi, and K.~M. Rabie, ``Physical layer security with {RF}
  energy harvesting in {AF} multi-antenna relaying networks,'' \emph{IEEE
  Trans. Commun.}, vol.~64, no.~7, pp. 3025--3038, Jul. 2016.

\bibitem{Li2016Secure}
Q.~Li, Q.~Zhang, and J.~Qin, ``Secure relay beamforming for {SWIPT} in
  amplify-and-forward two-way relay networks,'' \emph{IEEE Trans. Veh.
  Technol.}, vol.~65, no.~11, pp. 9006--9019, Nov. 2016.

\bibitem{Xing2015To}
H.~Xing, K.~K. Wong, Z.~Chu, and A.~Nallanathan, ``To harvest and jam: A
  paradigm of self-sustaining friendly jammers for secure {AF} relaying,''
  \emph{IEEE Trans. Signal Process.}, vol.~63, no.~24, pp. 6616--6631, Dec.
  2015.

\bibitem{Feng2017Robust}
Y.~Feng, Z.~Yang, W.~P. Zhu, Q.~Li, and B.~Lv, ``Robust cooperative secure
  beamforming for simultaneous wireless information and power transfer in
  amplify-and-forward relay networks,'' \emph{IEEE Trans. Veh. Technol.},
  vol.~66, no.~3, pp. 2354--2366, Mar. 2017.

\bibitem{Li2017Secure}
B.~Li, Z.~Fei, Z.~Chu, and Y.~Zhang, ``Secure transmission for heterogeneous
  cellular networks with wireless information and power transfer,'' \emph{IEEE
  Syst. J.}, vol.~PP, no.~99, pp. 1--12, 2017.

\bibitem{Niu2017Joint}
H.~Niu, B.~Zhang, D.~Guo, and Y.~Huang, ``Joint robust design for secure {AF}
  relay networks with {SWIPT},'' \emph{IEEE Access}, vol.~5, pp. 9369--9377,
  2017.

\bibitem{Li2016Robust}
B.~Li, Z.~Fei, and H.~Chen, ``Robust artificial noise-aided secure beamforming
  in wireless-powered non-regenerative relay networks,'' \emph{IEEE Access},
  vol.~4, pp. 7921--7929, 2016.

\bibitem{MP2009Directional}
M.~P. Daly and J.~T. Bernhard, ``Directional modulation technique for phased
  arrays,'' \emph{IEEE Trans. Antennas Propag.}, vol.~57, no.~9, pp.
  2633--2640, Sep 2009.

\bibitem{Yuan2014A}
Y.~Ding and V.~Fusco, ``A vector approach for the analysis and synthesis of
  directional modulation transmitters,'' \emph{IEEE Trans. Antennas Propag.},
  vol.~62, no.~1, pp. 361--370, Jan. 2014.

\bibitem{Ding2015Orthogonal}
------, ``Orthogonal vector approach for synthesis of multi-beam directional
  modulation transmitters,'' \emph{IEEE Antennas Wireless Propag. Lett.},
  vol.~14, pp. 1330--1333, 2015.

\bibitem{Hu2016Robust}
J.~Hu, F.~Shu, and J.~Li, ``Robust synthesis method for secure directional
  modulation with imperfect direction angle,'' \emph{IEEE Commun. Lett.},
  vol.~20, no.~6, pp. 1084--1087, Jun. 2016.

\bibitem{Shu2016Robust}
F.~Shu, X.~Wu, J.~Li, R.~Chen, and B.~Vucetic, ``Robust synthesis scheme for
  secure multi-beam directional modulation in broadcasting systems,''
  \emph{IEEE Access}, vol.~4, no.~99, pp. 6614--6623, 2016.

\bibitem{Liu2014Secrecy}
L.~Liu, R.~Zhang, and K.~C. Chua, ``Secrecy wireless information and power
  transfer with {MISO} beamforming,'' \emph{IEEE Trans. Signal Process.},
  vol.~62, no.~7, pp. 1850--1863, Apr. 2014.

\bibitem{Improving_Dong}
L.~Dong, Z.~Han, A.~P. Petropulu, and H.~V. Poor, ``Improving wireless physical
  layer security via cooperating relays,'' \emph{IEEE Trans. Signal Process.},
  vol.~58, no.~3, pp. 1875--1888, Mar. 2010.

\bibitem{Mukherjee2010Robust}
A.~Mukherjee and A.~L. Swindlehurst, ``Robust beamforming for security in
  {MIMO} wiretap channels with imperfect {CSI},'' \emph{IEEE Trans. Signal
  Process.}, vol.~59, no.~1, pp. 351--361, Jan. 2011.

\bibitem{Alam2017Asymptotic}
A.~M. Alam, P.~Mary, J.~Y. Baudais, and X.~Lagrange, ``Asymptotic analysis of
  area spectral efficiency and energy efficiency in {PPP} networks with {SLNR}
  precoder,'' \emph{IEEE Trans. Commun.}, vol.~65, no.~7, pp. 3172--3185, Jul.
  2017.

\bibitem{Charnes}
A.~Charnes and W.~W. Cooper, ``Programming with linear fractional
  functionals,'' \emph{Naval Research Logistics}, vol.~9, no. 3-4, pp.
  181--186, 1962.

\bibitem{Boyd}
S.~Boyd and L.~Vandenberghe, \emph{Convex Optimization}.\hskip 1em plus 0.5em
  minus 0.4em\relax Cambridge U.K.: Cambridge Univ. Press, 2004.

\bibitem{Li2015}
Q.~Li, Y.~Yang, W.~K. Ma, M.~Lin, J.~Ge, and J.~Lin, ``Robust cooperative
  beamforming and artificial noise design for physical-layer secrecy in {AF}
  multi-antenna multi-relay networks,'' \emph{IEEE Trans. Signal Process.},
  vol.~63, no.~1, pp. 206--220, Jan. 2015.

\bibitem{Shu2016Adaptive}
F.~Shu, J.~Tong, X.~You, G.~U. Chen, and W.~U. Jiajun, ``Adaptive robust
  {M}ax-{SLNR} precoder for {MU-MIMO-OFDM} systems with imperfect {CSI},''
  \emph{Science China Information Sciences}, vol.~59, no.~6, pp. 1--14, Jul.
  2016.

\bibitem{SecreyrateVPoor}
A.~A. Nasir, H.~D. Tuan, T.~Q. Duong, and H.~V. Poor, ``Secrecy rate
  beamforming for multicell networks with information and energy harvesting,''
  \emph{IEEE Trans. Signal Process.}, vol.~65, no.~3, pp. 677--689, Feb 2017.

\bibitem{Zappone2017Globally}
A.~Zappone, E.~Bjornson, L.~Sanguinetti, and E.~Jorswieck, ``Globally optimal
  energy-efficient power control and receiver design in wireless networks,''
  \emph{IEEE Trans. Signal Process.}, vol.~65, no.~11, pp. 2844--2859, Jun.
  2017.

\bibitem{KY2014Outage}
K.~Y. Wang, A.~M.~C. So, T.~H. Chang, W.~K. Ma, and C.~Y. Chi, ``Outage
  constrained robust transmit optimization for multiuser miso downlinks:
  Tractable approximations by conic optimization,'' \emph{IEEE Trans. Signal
  Process.}, vol.~62, no.~21, pp. 5690--5705, Nov 2014.

\bibitem{Gradshteyn2007}
I.~S. Gradshteyn and I.~M. Ryzhik, \emph{Table of Integrals, Series, and
  Products, Seventh Edition}.\hskip 1em plus 0.5em minus 0.4em\relax San Diego,
  CA, USA: Academic, 2007.

\bibitem{Golub1996Matrix}
G.~H. Golub and C.~F. Van~Loan, \emph{Matrix computations}.\hskip 1em plus
  0.5em minus 0.4em\relax John Hopkins University Press, 1996.

\end{thebibliography}


\ifCLASSOPTIONcaptionsoff
  \newpage
\fi




\end{document}